\documentclass[journal=ancac3,manuscript=article]{achemso}
\pdfoutput=1
\usepackage{graphicx}
\usepackage{amsmath}
\usepackage{amsfonts}
\setkeys{acs}{etalmode=truncate,maxauthors=30}
\SectionNumbersOn

\title{Current-Induced Torques with Dresselhaus Symmetry Due to Resistance Anisotropy in 2D Materials}

\author{Gregory M. Stiehl}
\affiliation{Department of Physics, Cornell University, Ithaca, NY 14853, USA}
\author{David MacNeill}
\affiliation{Department of Physics, Cornell University, Ithaca, NY 14853, USA}
\author{Nikhil Sivadas}
\affiliation{School of Applied and Engineering Physics, Cornell University, Ithaca, NY 14853, USA}
\author{Ismail El Baggari}
\affiliation{School of Applied and Engineering Physics, Cornell University, Ithaca, NY 14853, USA}
\author{Marcos H. D. Guimar\~{a}es}
\affiliation{Department of Physics, Cornell University, Ithaca, NY 14853, USA}
\alsoaffiliation{Kavli Institute at Cornell, Cornell University, Ithaca, NY 14853, USA}
\author{Neal D. Reynolds}
\affiliation{Department of Physics, Cornell University, Ithaca, NY 14853, USA}
\author{Lena F. Kourkoutis}
\affiliation{School of Applied and Engineering Physics, Cornell University, Ithaca, NY 14853, USA}
\alsoaffiliation{Kavli Institute at Cornell, Cornell University, Ithaca, NY 14853, USA}
\author{Craig J. Fennie}
\affiliation{School of Applied and Engineering Physics, Cornell University, Ithaca, NY 14853, USA}
\author{Robert A. Buhrman}
\affiliation{School of Applied and Engineering Physics, Cornell University, Ithaca, NY 14853, USA}
\author{Daniel C. Ralph}
\affiliation{Department of Physics, Cornell University, Ithaca, NY 14853, USA}
\alsoaffiliation{Kavli Institute at Cornell, Cornell University, Ithaca, NY 14853, USA}
 \email{dcr14@cornell.edu}
\date{{\small \today}}
\keywords{transition metal dichalcogenides, spin-orbit torque, spin-torque, crystal symmetry, van der Waals materials, 2D materials}
\begin{document}
\newpage
\begin{abstract}
We report measurements of current-induced torques in heterostructures of Permalloy (Py) with TaTe$_2$, a transition-metal dichalcogenide (TMD) material possessing low crystal symmetry, and observe a torque component with Dresselhaus symmetry. We suggest that the dominant mechanism for this Dresselhaus component is not a spin-orbit torque, but rather the Oersted field arising from a component of current that flows perpendicular to the applied voltage due to resistance anisotropy within the TaTe$_2$.  This type of transverse current is not present in wires made from a single uniform layer of a material with resistance anisotropy, but will result whenever a material with resistance anisotropy is integrated into a heterostructure with materials having different resistivities, thereby producing a spatially non-uniform pattern of current flow. This effect will therefore influence measurements in a wide variety of heterostructures incorporating 2D TMD materials and other materials with low crystal symmetries. 
\end{abstract}

\maketitle

\begin{figure*}[!t]
	\centering
	\includegraphics[width=14 cm]{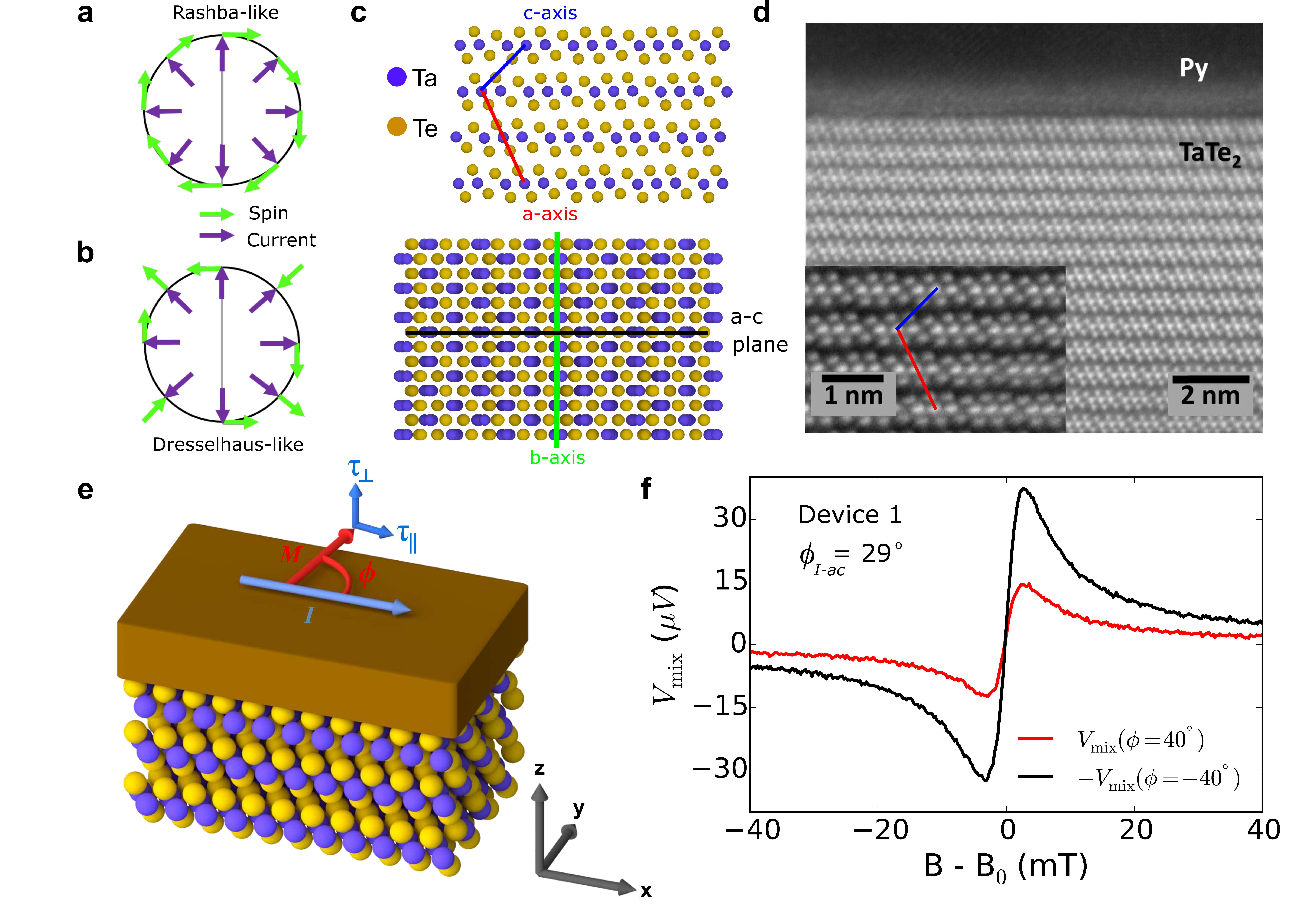}
	\caption[Sample geometries and ST-FMR mixing voltage.]{(a) Rashba-like and (b) Dresselhaus-like net spin (or field) polarizations, where the spin (green arrow) is generated in response to an applied current (purple arrow). The grey line represents a mirror plane. (c) TaTe$_2$ crystal structure looking down the b-axis (top) and the exfoliation plane (bottom). The yellow spheres represent Te atoms and the purple spheres represent Ta atoms. (d) A cross-sectional HAADF-STEM image of a TaTe$_2$/Permalloy device showing high crystallinity except for a region at the TaTe$_2$/Py interface approximately one TaTe$_2$ layer thick.  In all other layers, the trimerization associated with the low-symmetry room-temperature TaTe$_2$ crystal structure is clearly visible. (Inset) a HAADF-STEM image of the same device with higher magnification, clearly showing the low-symmetry structure. (e) Schematic of the TaTe$_2$/Permalloy sample geometry. The x-axis is defined to be parallel to the applied electric field and the z-axis is perpendicular to the sample plane. (f) ST-FMR resonances for a TaTe$_2$ (19.7 nm) / Py (6 nm) device (Device 1) with the magnetization oriented at 40$^{\circ}$ and -40$^{\circ}$ degrees with respect to the current direction. The applied magnetic field, $B$, is normalized by the resonance field, $B_0$, to account for a small shift in the resonance due to an in-plane uniaxial anisotropy in the Permalloy. The applied microwave power is 2 dBm at a frequency of 9 GHz.}
	\label{TaTe2fig1}
\end{figure*}

\newpage
Current-induced spin-orbit torques are a promising method for efficiently manipulating magnetic devices\ \cite{Brataas2012}.  Understanding the mechanisms by which the directions of these torques can be manipulated, for example by using crystal symmetries, is important for optimizing them for applications.  To date, all observations of spin-orbit torques from centrosymmetric materials –-- generated through either spin Hall\ \cite{Liu2011,Liu555}, Rashba-Edelstein\ \cite{Miron2010,Sanchez2013}, topological spin-momentum locking\ \cite{Mellnik2014,Fan2014}, or other spin-orbit effects\ \cite{StilesPRBForm,StilesPRBPhen} –-- can be described as corresponding to a Rashba-like symmetry (Fig.\ \ref{TaTe2fig1}a). That is, the generated field or spin is perpendicular to the applied current and lies within the sample plane. Torques corresponding to a more general spin symmetry have been observed only in non-centrosymmetric systems, such as torques resulting from the out-of-plane spins in WTe$_2$\ \cite{MacNeill2016,MacNeill2017}, or torques corresponding to a Dresselhaus-like spin polarization (Fig.\ \ref{TaTe2fig1}b) observed in GaMnAs\ \cite{Fang2011,Kurebayashi2014}, GaAs/Fe heterostructures\ \cite{Skinner2015,Chen2016} and NiMnSb\ \cite{Ciccarelli2016}. Here, we analyze current-induced torques in heterostructures of Permalloy (Py = Ni$_{81}$Fe$_{19}$) with the low-symmetry material TaTe$_2$, a centrosymmetric transition-metal dichalcogenide (TMD). To our surprise, the heterostructures exhibit a component of field-like torque for which the dependence on the angle of applied current relative to the crystalline axes reflects a Dresselhaus symmetry, despite the fact that TaTe$_2$ is inversion symmetric. We suggest that in TaTe$_2$/Py this torque does not originate from a  spin-orbit mechanism. Instead, it likely arises from resistance anisotropy within the plane of the TaTe$_2$ layers, which in TaTe$_2$/Py heterostructures can cause current flow non-collinear with the applied electric field, leading to an Oersted field that mimics a Dresselhaus symmetry. This effect will modify the form of current-induced torques produced by any low-symmetry source material, and might be used beneficially to engineer the direction of the Oersted torque to assist switching in memory devices \cite{Aradhya2016}. 

The transverse current flows we analyze will not occur in wires made from a single layer of a uniform material having a resistance anisotropy.  In that case, in order to satisfy the boundary condition that there be zero transverse current at the edge of the wire, a transverse voltage will be generated to cancel any transverse component of applied current, and the Oersted field generated by the applied current will not possess any unusual symmetry.  However, if one adds one or more layers with a different resistivity, the additional layers will provide a return path to allow a spatially non-uniform loop of transverse current.  This is a common situation in samples incorporating low-symmetry 2D crystalline materials including many TMDs. These spatially non-uniform current flows therefore have the potential to influence many types of experiments involving heterostructures of 2D materials. 

\section*{Results}
TaTe$_2$ at room temperature has a monoclinic (1T') crystal structure with a centrosymmetric space group C2/m (\# 12)\ \cite{Brown1966,SORGEL2006}. When integrated into a heterostructure with Py, only a single structural symmetry remains: a mirror plane perpendicular to the TMD layers. In TaTe$_2$, this mirror is within the a-c plane (Fig. \ref{TaTe2fig1}c). The low-symmetry crystal structure of TaTe$_2$ is clearly visible in the cross-sectional high-angle annular dark-field scanning transmission electron microscopy (HAADF-STEM) image of one of our devices (Fig. \ref{TaTe2fig1}d).  

\begin{figure}[!t]
	\centering
	\includegraphics[width=15 cm]{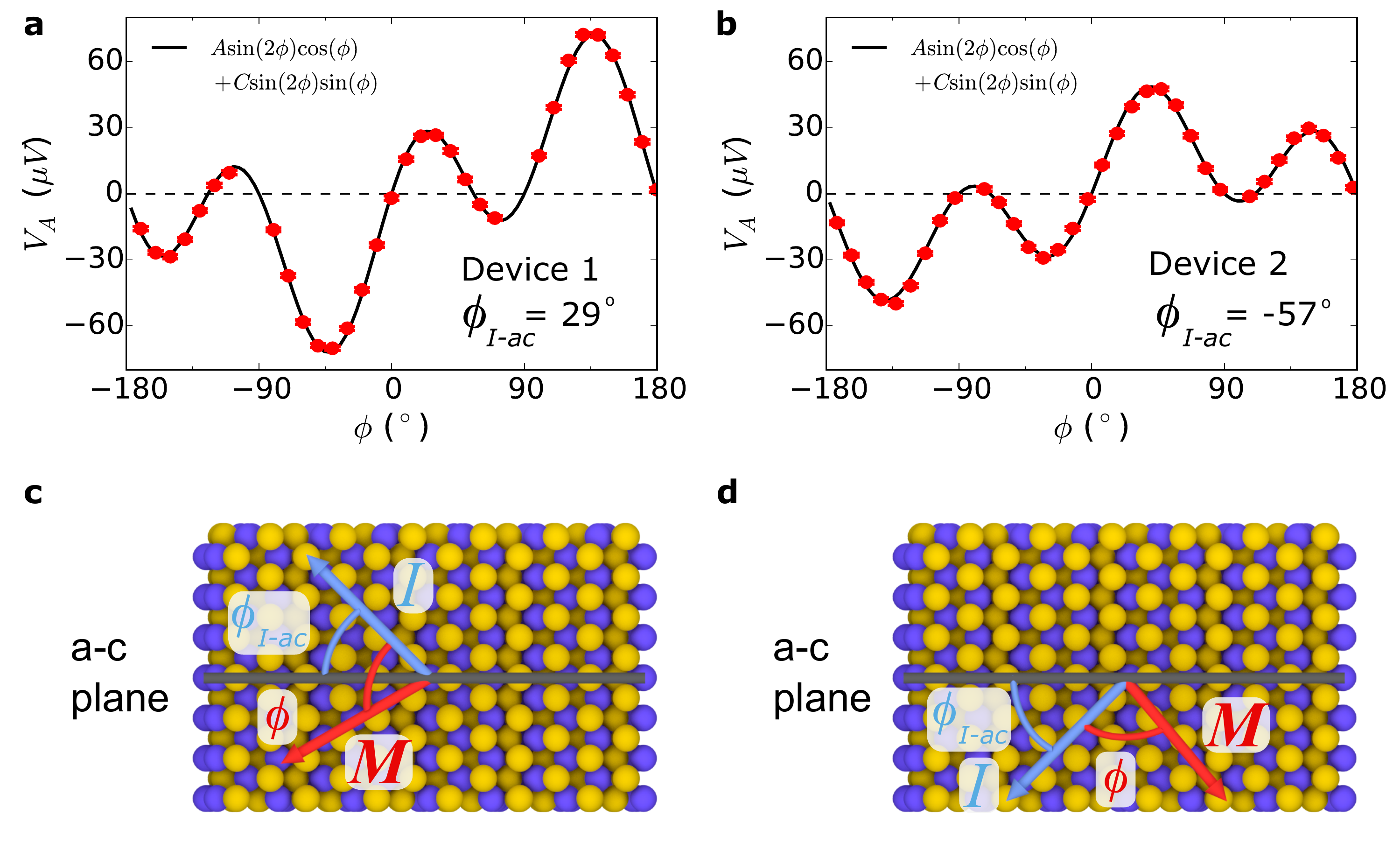}
	\caption[Angular dependence of out-of-plane torques in TaTe$_2$. ]{(a) and (b) Antisymmetric component of the ST-FMR resonance as a function of applied in-plane magnetic field angle for TaTe$_2$/Py Devices 1 and 2 respectively. The applied microwave power is 2 dBm at a frequency of 9 GHz. Device 1 is TaTe$_2$ (19.7 nm) / Py (6 nm) and Device 2 is TaTe$_2$ (8.8 nm) / Py (6 nm). The value of ${\phi _{I - ac}}$, the angle between the current and TaTe$_2$ a-c mirror plane, is 29$^{\circ}$ in panel (a) and -57$^{\circ}$ in panel (b). (c) and (d) depict a positive and negative $\phi_{I-ac}$ angle, respectively, corresponding to Devices 1 and 2, and $\phi$ is depicted as positive for both (c) and (d). }
	\label{TaTe2fig3}
\end{figure}

We characterize the angle ${\phi _{I - ac}}$ between the direction of the applied current and the mirror plane of the TMD in the finished devices by measurements of a magnetic easy axis in the Py induced by interaction with the TMD\ \cite{MacNeill2016,MacNeill2017,Haney2016}, in combination with polarized Raman spectroscopy and HAADF-STEM imaging (see Supporting Information). The Py equilibrium magnetization direction, $\hat{m}$, lies within the sample plane. As depicted in Fig. \ref{TaTe2fig1}e, when a current is applied to a TMD/Py heterostructure a current-induced torque acts on the magnetic moment. To measure the current-induced torques in our samples, we use two complementary measurement methods, a harmonic Hall technique\ \cite{MacNeill2017,Hayashi2014,Avci2014}, and spin-torque ferromagnetic resonance (ST-FMR)\ \cite{Mellnik2014,Liu2011,MacNeill2016}, with all measurements made at room temperature.  Both types of measurements gave consistent results.  The harmonic Hall measurements are detailed in the Supporting Information. 

In the ST-FMR measurements, an in-plane RF current (7-12 GHz) is applied to the sample which generates torques on the ferromagnet in phase with the current (Fig. \ref{TaTe2fig1}e).  An in-plane magnetic field is applied at an angle of $\phi $ relative to the applied current, and the magnitude of this field is swept through the ferromagnetic resonance condition. We measure a DC voltage arising from mixing between the RF current and resistance oscillations resulting from magnetization precession together with the anisotropic magnetoresistance (AMR) of the Py (Fig. \ref{TaTe2fig1}f). This mixing voltage, ${V_{{\rm{mix}}}}$, can be fitted as a function of magnetic field as a sum of symmetric and antisymmetric Lorentzians, where the amplitudes of these resonances (${V_S}$ and ${V_{\rm{A}}}$) allow independent measurements of the in-plane (${\vec \tau _\parallel }$) and out-of-plane (${\vec \tau _ \bot }$) spin-orbit torques respectively\ \cite{Liu2011,MacNeill2016}:
\begin{equation}
{V_{\rm{S}}} =  - \frac{{{I_{{\rm{RF}}}}}}{2}\left( {\frac{{dR}}{{d\phi }}} \right)\frac{1}{{{\alpha _{\rm{G}}}\gamma \left( {2{B_0} + {\mu _0}{M_{{\rm{eff}}}}} \right)}}{\tau _\parallel },\label{TaTe2Vs}
\end{equation}
\begin{equation}
{V_{\rm{A}}} =  - \frac{{{I_{{\rm{RF}}}}}}{2}\left( {\frac{{dR}}{{d\phi }}} \right)\frac{{\sqrt {1 + {\mu _0}{M_{{\rm{eff}}}}/{B_0}} }}{{{\alpha _{\rm{G}}}\gamma \left( {2{B_0} + {\mu _0}{M_{{\rm{eff}}}}} \right)}}{\tau _ \bot }. \label{TaTe2Va}
\end{equation}	 	
Here $R$ is the device resistance, $dR/d\phi $ is due to the AMR in the Py, ${\mu _0}{M_{eff}}$ is the out-of-plane demagnetization field, ${B_0}$ is the resonance field, ${I_{RF}}$ is the microwave current in the heterostructure, ${\alpha _G}$ is the Gilbert damping coefficient, and the equilibrium magnetization is saturated along the applied field direction. 

The magnitude of torques with a conventional Rashba-like symmetry, ${\vec \tau _\parallel } \propto \hat m \times (\hat m \times \hat y)$ and ${\vec \tau _ \bot } \propto \hat m \times \hat y$ for current in the $\hat x$ direction, are proportional to $\cos (\phi )$. Therefore, in the presence of only Rashba-like torques the magnitude of ${V_{mix}}$ is unchanged upon the operation $\phi  \to  - \phi $ but the sign is reversed (as $dR/d\phi  \propto \sin (2\phi )$ in Eqs. \ref{TaTe2Vs} and \ref{TaTe2Va}). Figure \ref{TaTe2fig1}f shows resonance curves in ${V_{mix}}$ as a function of applied in-plane field magnitude for $\phi  = {40^ \circ }$ (red) and $\phi  =  - {40^ \circ }$ (black, inverted), for one of our TaTe$_2$/Py devices (Device 1). The difference between the two ${V_{mix}}$ measurements shows a lack of $\phi  \to  - \phi $ symmetry in the observed torques and suggests the presence of a torque which does not arise entirely from a Rashba-like spin polarization.  For all of the TaTe$_2$/Py devices, the antisymmetric component of the ST-FMR resonance is by far the dominant contribution, so we will focus on ${\vec \tau _ \bot }$ here in the main text.  The symmetric ST-FMR component indicates only a weak in-plane antidamping torque with Rashba symmetry ${\vec \tau _\parallel } \propto \hat m \times (\hat m \times \hat y)$ and in some cases a small contribution $ \propto \hat m \times \hat z$ that is not consistent from sample to sample and might arise from strain\cite{Guimaraes2018Nano} (see Supporting Information).

Figure \ref{TaTe2fig3}a shows ${V_A}$ as a function of $\phi $ for TaTe$_2$/Py Device 1. The observed ${V_A}(\phi )$ clearly lacks $\phi  \to  - \phi $ symmetry and therefore cannot be described as arising solely from Rashba-like torques $ \propto \cos (\phi )$. Other symmetries are allowed, however, in low-symmetry samples such as TaTe$_2$/Py. Torques associated with Dresselhaus-like spin generation (Fig. \ref{TaTe2fig1}b) can contribute components ${\vec \tau _\parallel } \propto \hat m \times [\hat m \times [\cos (2{\phi _{I - ac}})\hat y \pm \sin (2{\phi _{I - ac}})\hat x]]$ and ${\vec \tau _ \bot } \propto \hat m \times [\cos (2{\phi _{I - ac}})\hat y \pm \sin (2{\phi _{I - ac}})\hat x]$ where $\hat x$ is the direction of applied current.  The parts of the Dresselhaus contributions proportional to $\hat m \times (\hat m \times \hat x)$ or $\hat m \times \hat x$ will give torque amplitudes $ \propto \sin (2{\phi _{I - ac}})\sin (\phi )$.  We will refer to any current-induced torque of this form as Dresselhaus-like, regardless of its microscopic origin. If we model the out-of-plane torques in our TaTe$_2$/Py heterostructures as a sum of Rashba-like and Dresselhaus-like terms, we can fit ${V_A}$ as: 
\begin{equation}
{V_A} = \sin (2\phi )\left[ {A\cos (\phi ) + C\sin (\phi )} \right],\label{VaPhiC}
\end{equation}
where the $\sin (2\phi )$ dependence comes from the AMR ($dR/d\phi $) in Eq. \ref{TaTe2Va}, and both A and C might depend on ${\phi _{I - ac}}$. We extract a value of $C/A =  - 0.69 \pm 0.01$ for Device 1, in which ${\phi _{I - ac}}$ is positive. In Fig. \ref{TaTe2fig3}b we show ${V_A}(\phi )$ for TaTe$_2$/Py Device 2, in which ${\phi _{I - ac}}$ is negative. Positive and negative values of ${\phi _{I - ac}}$ are as defined in Fig.\ \ref{TaTe2fig3}c and d respectively. In Device 2 the sign of the $\phi  \to  - \phi $ symmetry breaking is opposite that in Device 1, corresponding to an opposite sign $C/A = 0.38 \pm 0.01$. This is consistent with the expectation that in a Dresselhaus-like symmetry the component of spin or field along the current direction changes sign across the mirror plane (Fig. \ref{TaTe2fig1}b). We note that the observation of a torque $ \propto \hat m \times \hat x$ is distinct from our previously-published work on WTe$_2$/Py\ \cite{MacNeill2016,MacNeill2017}, in which we observed a different non-Rashba component of ${\vec \tau _ \bot } \propto \hat m \times (\hat m \times \hat z)$. A torque proportional to $\hat m \times (\hat m \times \hat z)$ for an in-plane magnetization amounts to adding a term constant in $\phi $ to Eq. \ref{VaPhiC} ($B$), such that ${\tau _ \bot } = A\cos (\phi ) + B  + C\sin (\phi )$. We observe no out-of-plane antidamping torque in our TaTe$_2$/Py devices within experimental uncertainty, even though such a torque is symmetry-allowed in the heterostructure. 

\begin{figure*}[tbph!]
	\centering
	\includegraphics[width=9 cm]{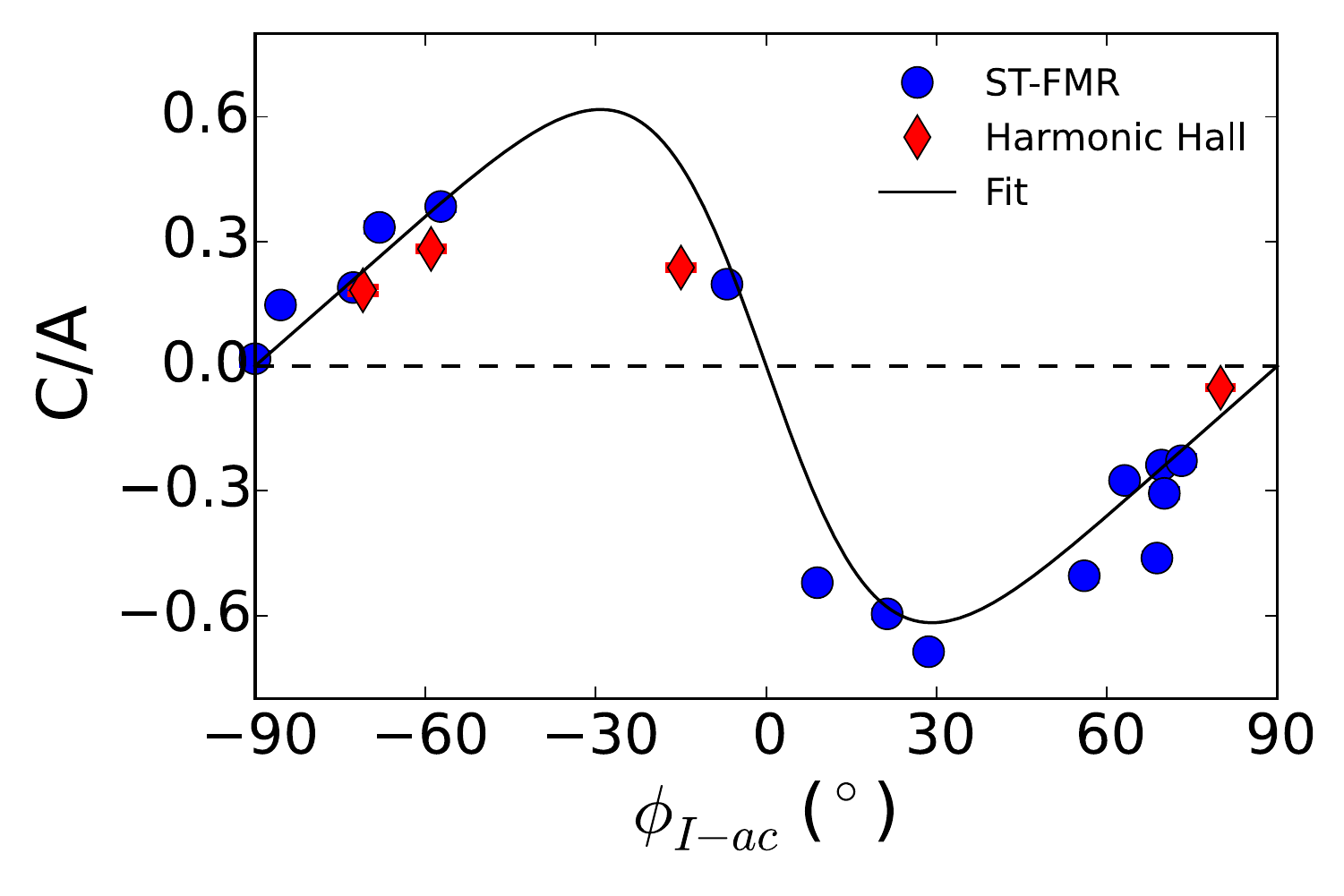}
	\caption{Ratio of torques $ \propto \hat m \times \hat x$  to the torques $ \propto \hat m \times \hat y$, C/A, as a function of the angle between the applied current and the TaTe$_2$ a-c mirror plane for devices studied by either ST-FMR (blue circles) or second-harmonic Hall measurements (red diamonds). The fit is discussed in the Supporting Information.}
	\label{TaTe2fig5}
\end{figure*}

We have performed torque measurements on 19 different TaTe$_2$/Py devices (4 second-harmonic Hall devices and 15 ST-FMR devices), all with distinct values of ${\phi _{I - ac}}$ and TaTe$_2$ thicknesses, ${t_{TMD}}$. Figure\ \ref{TaTe2fig5} shows extracted values of $C/A$ as a function of ${\phi _{I - ac}}$ for both types of samples.  The measurements are in good agreement with the dependence on ${\phi _{I - ac}}$ expected for a field or spin polarization with Dresselhaus symmetry (Fig.\ \ref{TaTe2fig1}b): $C/A$ goes to zero when the current is applied either along or perpendicular to a mirror plane (${\phi _{I - ac}}=0^{\mathrm{o}}$, 90$^{\mathrm{o}}$, and 180$^{\mathrm{o}}$), and changes sign as ${\phi _{I - ac}}$ crosses the TaTe$_2$ mirror plane (${\phi _{I - ac}} = {0^ \circ }$). Details for each device are given in the Supporting Information.

To obtain a more quantitative estimate for the strength of the Dresselhaus-like torques, we take into account that a Dresselhaus torque does not point exclusively in the direction ${\vec \tau _ \bot } \propto \hat m \times \hat x$, but depending on the value of ${\phi _{I - ac}}$ it can also have a component in the perpendicular direction that can add to or subtract from a component with Rashba symmetry (see Fig. 1a,b):
	\[\tau _ \bot ^C = C\sin (\phi ) = [D\sin (2{\phi _{I - ac}})]\sin (\phi ),\] 	
	\[\tau _ \bot ^A = A\cos (\phi ) = [R + D\cos (2{\phi _{I - ac}})]\cos (\phi ),\] 
where $R$ is the component of $\cos (\phi )$ torques arising from a Rashba-like symmetry, and $D$ for Dresselhaus-like.  The fit lines shown in Fig.\ \ref{TaTe2fig5} for TaTe$_2$/Py corresponds to a value $D/R= -0.51\pm 0.03$ (see Supporting Information).

\section*{Discussion}
We now turn to consideration of the microscopic mechanism that generates current-induced torques with Dresselhaus-like symmetry in our system.  TaTe$_2$ cannot generate a torque with a Dresselhaus symmetry through the mechanism present in GaMnAs\ \cite{Fang2011,Kurebayashi2014} and NiMnSb\ \cite{Ciccarelli2016} (a bulk inverse spin Galvanic effect), since inversion symmetry is intact in the TaTe$_2$ bulk. An {\it interfacial} spin-orbit-torque mechanism is symmetry-allowed, but this would imply a dependence on the TaTe$_2$ thickness that is inconsistent with our measurements (see Supporting Information). In fact, the measured thickness dependence of the Dresselhaus-like term tracks that of the conventional Rashba-symmetry field-like torque (see Supporting Information), suggesting that both arise from a current-generated Oersted field (see Supporting Information). Furthermore, preliminary first-principles modeling of interfacial spin-orbit torques $via$ a ``hidden spin-polarization"\ mechanism \cite{ZhangNatPhys2014,Zelezny2017,Wadley587} (where symmetry mandates that any current-induced spin-polarization on one Te surface must be equal but opposite to the spin-polarization on the opposing Te surface within a single TaTe$_2$ layer) suggests that this effect is small in our system. Finally, except for the conventional Oersted torque and the Dresselhaus-like torque, our measurements indicate that all other components of current-induced torque in TaTe$_2$/Py are small or zero, including the in-plane antidamping torque with Rashba symmetry, $ \propto \hat m \times (\hat m \times \hat y)$, that is usually dominant in spin-orbit systems. We conclude that direct spin-orbit torques are simply weak in this system (which is likely the reason that TaTe$_2$ does not exhibit an out-of-plane antidamping torque $ \propto \hat m \times (\hat m \times \hat z)$ even though it is symmetry allowed).  

\begin{figure*}[!t]
\centering
\includegraphics[width=14 cm]{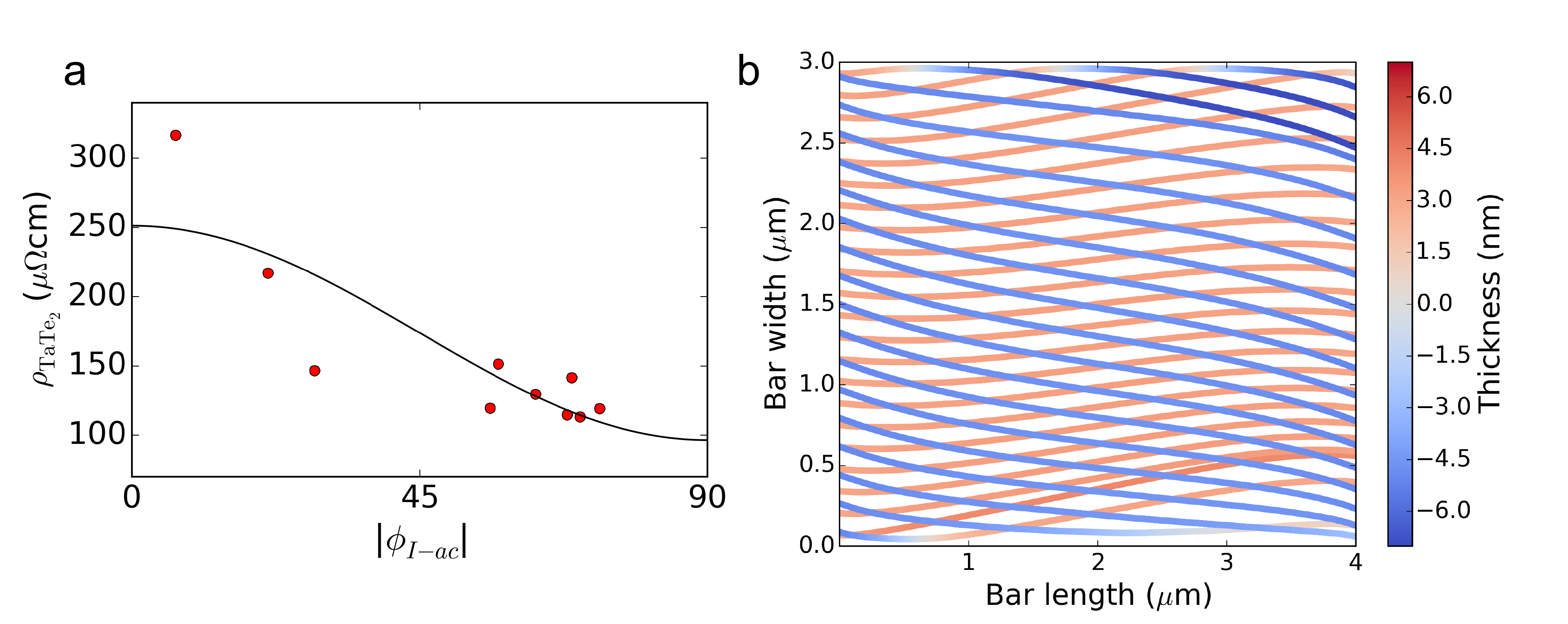}
    \caption[Extracted resistivities as a function of crystal axis and modeled current streamlines]{a) Measurement of the TaTe$_2$ resistivity for 10 of our devices as a function of $|{\phi _{I - ac}}|$, extracted from the two-point resistance. (b) Simulated current paths for a TaTe$_2$(10 nm)/Py(6 nm) bar of length 4 $\mu m$ and width 3 $\mu m$ with ${\phi _{I - ac}} = 45^\circ $  for a constant voltage applied across the length of the bar. The color map shows the height of the current streamline, with positive values in the Py layer and negative values in the TaTe$_2$ layer. }
    \label{TaTe2fig6}
\end{figure*}

We suggest, instead of a spin-orbit-torque mechanism, that the Dresselhaus-like torque arises from in-plane resistivity anisotropy within the TMD. TaTe$_2$, along with other low symmetry TMDs such as WTe$_2$ and 1T'-MoTe$_2$, exhibits significant resistance anisotropy. We show the extracted resistivity of TaTe$_2$ from our devices as a function of $|\phi_{I-ac}|$ in Fig\ \ref{TaTe2fig6}a (where we have removed contributions from the Py layer and contact resistance as outlined in the Supporting Information). The extracted in-plane resistivity anisotropy is 2.6$\pm$0.6. When an electric potential is applied away from one of the principal axes in a material with anisotropic resistivity, the electric field and the current are no longer collinear, $i.e.$ for a potential along the sample bar the generated current may be tilted. In a bar consisting of just one material, say TaTe$_2$, the boundary conditions force the transverse current at the edges of the bar to be zero. However, in a heterostructure with Py, the transverse component of current in the TMD will turn into the Py to establish a return current flowing in the reverse transverse direction and result in a circulating transverse current loop. The Oersted field generated by this current loop naturally produces a field-like torque on the Py layer that mimics Dresselhaus symmetry (${\vec \tau _ \bot } \propto \sin (2{\phi _{I - ac}})\hat m \times \hat x$), in addition to the standard Oersted torque with a Rashba symmetry from the projection of current flowing along the bar. We have modeled the current pathways and associated Oersted fields in our TaTe$_2$/Py heterostructures through the finite element analysis software package COMSOL. Figure\ \ref{TaTe2fig6}b shows the simulated current path for a constant voltage applied across the length of a TaTe$_2$/Py heterostructure (length 4 $\mu$m and width of 3 $\mu$m) with an in-plane resistivity anisotropy ratio of 2.6 in the TaTe$_2$. We consider the case that the principle axes of the TaTe$_2$ crystal are tilted at a 45$^\circ$ angle from the length of the bar (${\phi _{I - ac}} = 45^\circ $). The blue streamlines show the current within the TaTe$_2$, and the red streamlines show the current flowing within the Py. 

By taking the ratio of the integrated total current within the TaTe$_2$ layer going in the y-direction (along the width of the bar) to that in the x-direction (along the length of the bar) we can estimate a value of C/A due to the Oersted field generated by tilted currents.  The result has the same dependence on ${\phi _{I - ac}}$ as measured for the TaTe$_2$/Py heterostructures and the correct overall sign of C/A vs. ${\phi _{I - ac}}$ for TaTe$_2$/Py.  For an anisotropy ratio of 2.6 we estimate a ratio of $C/A \sim 0.32$, within a factor of 2 of the result found in experiment.  The quantitative difference might be explained by an underestimate of the resistivity anisotropy in the TaTe$_2$ or by spatial non-uniformity in the resistivity of the Py layer as a function of thickness.  In the HAADF-STEM imaging (Fig. 1d) we observe some intermixing at the TaTe$_2$/Py interface in a region of approximately one TaTe$_2$ layer thickness.  If this disorder causes increased scattering in the Py near the TaTe$_2$/Py interface, the average effective resistivity of the Py would be higher below the midplane of the Py layer than above. This would cause the return current of the transverse current loop flowing in the Py to add to the Oersted field from the transverse current flowing in the TaTe$_2$, while the longitudinal current in the Py at the same time subtracts from the standard Oersted field produced by the longitudinal current in the TaTe$_2$. This has the overall effect of increasing $C$ and decreasing $A$, giving an enhanced value of $C/A$.  

Previously, our group has studied current-generated torques from another low-symmetry TMD, WTe$_2$, finding a different unusual component of spin-orbit torque -- an out-of-plane antidamping torque -- consistent with the WTe$_2$ crystal symmetries. Like TaTe$_2$, WTe$_2$ has in-plane resistivitity anisotropy, on the order of 2, so one should expect a field-like torque component with Dresselhaus-like symmetry there as well. In our previous work on WTe$_2$/Py samples\ \cite{MacNeill2016,MacNeill2017}, we studied primarily devices with current applied near high symmetry directions (${\phi _{I - bc}}=0^\circ$ and $\pm90^\circ$) and the Dresselhaus contribution was sufficiently small that we did not make note of it.  Nevertheless, measurements of WTe$_2$/Py devices at intermediate angles ${\phi _{I - bc}}$ allow a clear separation of the different torque components based on their dependence on $\phi $, and we do indeed observe a Dresselhaus-like component with $D/R = -0.13 \pm 0.02$ (see Supporting Information). The addition of this Dresselhaus-like torque in the analysis of our WTe$_2$/Py samples does not affect any of our previous conclusions about the strength of the out-of-plane antidamping torque.

\section*{Conclusions}	
In summary, we have measured current-induced torques with Dresselhaus-like symmetry in both TaTe$_2$/Py and WTe$_2$/Py heterostructures. We explain this torque component not by a direct spin-orbit-torque mechanism, but rather as due to the Oersted field generated by a component of current transverse to the applied voltage.  The transverse current arises from in-plane resistivity anisotropy of TaTe$_2$ and WTe$_2$ that generates spatially non-uniform current flows within the heterostructures.  This interesting effect will be present quite generally in heterostructures containing low-symmetry materials with in-plane resistivity anisotropy. It must  be taken into account when analyzing the angular dependence of spin-orbit torques in these systems, and when engineering low-symmetry materials to produce spin-orbit torques. It will also affect all other types of transport measurements on heterostructures containing 2D materials with resistance anisotropy whenever the applied voltage is not aligned with a symmetry axis.

\section*{Methods}
{\bf Sample Fabrication:} To fabricate our samples we exfoliate TaTe$_2$ from bulk crystals (supplied by HQ graphene) onto high resistivity silicon / silicon oxide wafers using the scotch tape method, where the final step of exfoliation is carried out in the load lock of our sputtering system under high vacuum ($<$ 10$^{-6}$ torr). Without breaking vacuum, we then deposit 6 nm of the ferromagnet permalloy (Py = Ni$_{81}$Fe$_{19}$) by grazing angle sputtering to minimize damage to the TaTe$_2$ surface (Fig.\ \ref{TaTe2fig1}d) in an Ar pressure of 4 mtorr. We use a deposition rate below 0.2 angstroms/second, with the substrate rotating at greater than 10 revolutions per minute. To prevent oxidation of the ferromagnet we cap the heterostructure with 2 nm of Al, which is oxidized upon exposure to atmosphere. Flakes for further processing are selected {\it ex situ} using optical and atomic force microscopy. All devices are positioned so that the active region is atomically flat, with an RMS surface roughness below 300 pm and no monolayer steps. The devices are patterned using e-beam lithography into either a microwave-frequency-compatible ground-signal-ground geometry for resonant measurements, or Hall bars for low-frequency (kHz) second-harmonic Hall measurements, with pattern transfer by Ar ion milling with SiO$_2$ used as an etch mask. The etched devices are protected by subsequent sputter coating of SiO$_2$. Electrical contacts, Ti (5 nm) / Pt (75 nm), are defined through a lift-off process.

\begin{acknowledgement}
The primary support for this project including support for G.M.S., who performed the sample fabrication, electronic and Raman spectroscopy measurements, and data analysis, came from the US Department of Energy (DE-SC0017671). G.M.S.\ wrote the manuscript with D.C.R.  D.M.\ assisted with  device fabrication and data analysis with support from the National Science Foundation (NSF) (DMR-1708499). N.S.\ contributed ab-initio modeling and discussion with support from the NSF through the Platform for the Accelerated Realization, Analysis, and Discovery of Interface Materials (PARADIM) (DMR-1539918) and the Cornell University Center for Advanced Computing at Cornell University. I.E.B.\ performed the electron microscopy under support from the NSF through PARADIM as part of the Materials for Innovation Platform Program. M.H.D.G. contributed to the data analysis and Raman measurements with support from the Kavli Institute at Cornell for Nanoscale Science and the Netherlands Organization for Scientific Research (NWO Rubicon 680-50-1311). N.D.R. contributed experimental assistance with support by the NSF through the Cornell Center for Materials Research (CCMR) (DMR-1719875). L.F.K., C.F., R.A.B., and D.C.R.\ supervised the research.  All authors contributed to the final version of the manuscript. Sample fabrication was performed in the CCMR shared facilities and at the Cornell Nanoscale Science \& Technology Facility, part of the National Nanotechnology Coordinated Infrastructure, which is supported by the NSF (ECCS-1542081). The FEI Titan Themis 300 was acquired through Grant NSF-MRI-1429155, with additional support from Cornell University, the Weill Institute, and the Kavli Institute at Cornell. 
\end{acknowledgement}

\section*{Supporting Information}
\setcounter{section}{0}
\setcounter{equation}{0}
\setcounter{figure}{0}
\renewcommand{\theequation}{S\arabic{equation}}
\renewcommand{\thefigure}{S\arabic{figure}}
\renewcommand{\thetable}{S\arabic{table}}
\section{Second harmonic Hall measurements}\label{SecHarmHallSec}

\begin{figure}[!th]
\centering
\includegraphics[width=7 cm]{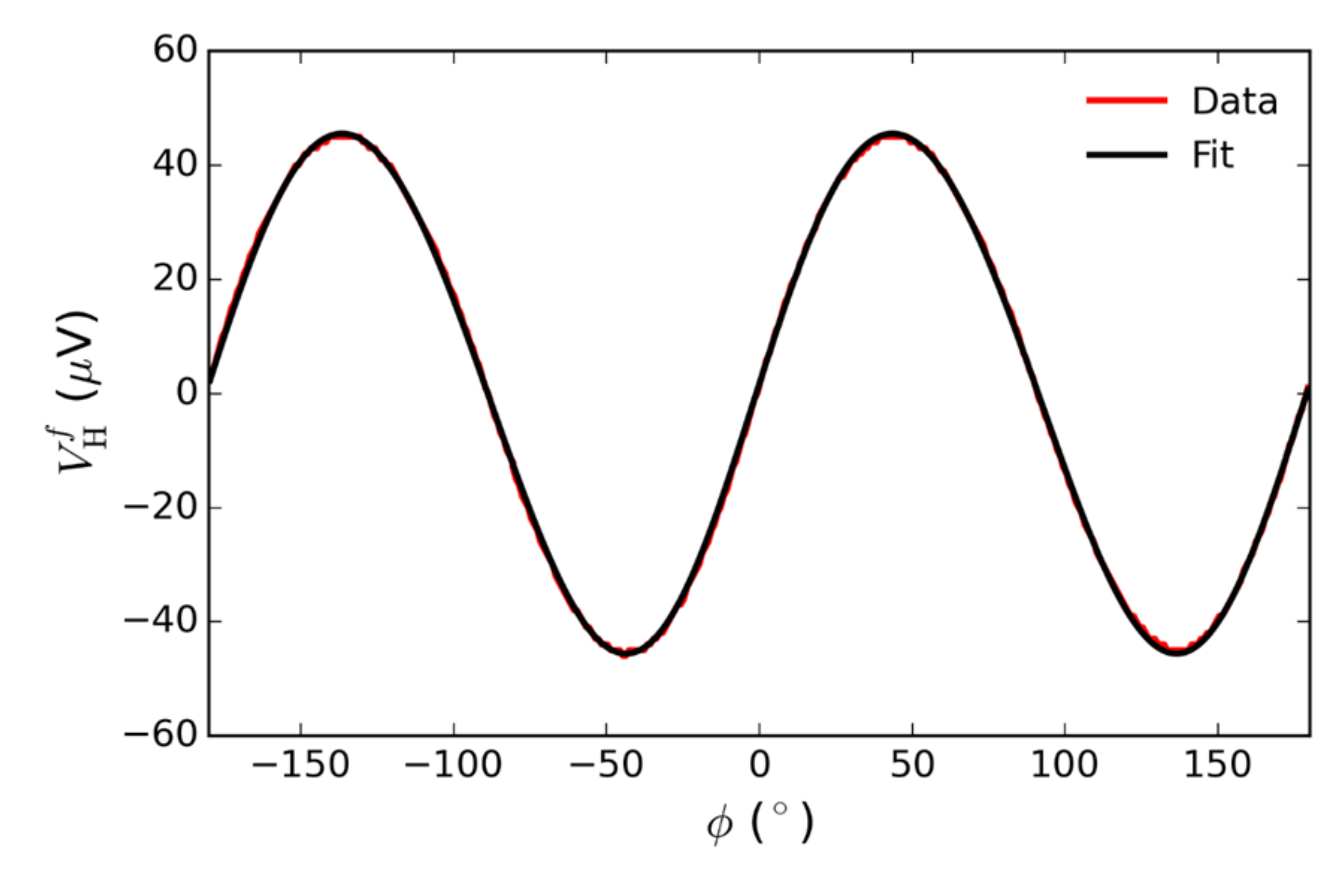}
    \caption[ First harmonic Hall measurement of a TaTe$_2$/Py device]{First harmonic Hall signal as a function of the angle of an in-plane magnetic field (red) and fit (black) in Device 16.  The magnitude of the magnetic field is 20 mT and a 300 mV RMS excitation is applied across a series connection of the 480 $\Omega$ sample and a 50 $\Omega $ bias resistor. }
    \label{TaTe2firstharm}
\end{figure}

Hall bars are fabricated using the same process as our ST-FMR devices, and have a length and width as specified in Table\ \ref{tab:TaTe2}; the width of the voltage probes used for the Hall measurements are scaled by a ratio of 0.375 times the width of the bar for each device. The active region of the Hall bar has a uniform TaTe$_2$ thickness, with no monolayer steps as measured by atomic force microscopy. We apply a voltage $V\left( t \right) = {V_0}\cos \left( {2\pi ft} \right)$ across the sample and a 50 $\Omega $ bias resistor in series at a frequency f=1.3 kHz, where ${V_0}$ = 300 mV or 200 mV root mean square (RMS) for bars of width 3 $\mu m$ and 2 $\mu m$ respectively. 
The first ($V_{\rm{H}}^f$) and second ($V_{\rm{H}}^{2f}$) harmonic of the Hall signals are measured simultaneously as a function of applied magnetic field angle, where the magnitude of the applied field is held constant throughout a given measurement (ranging from 0.01 to 0.1 T). The current through the Hall bar is measured separately under the same experimental conditions as the Hall measurement. Figure\ \ref{TaTe2firstharm} shows the first harmonic Hall signal as a function of applied magnetic field angle, $\phi $, and is fit using the equation: 
	\[V_{\rm{H}}^f = {I_0}{R_{PHE}}\sin (2{\phi _M}),\] 
where ${I_0}$ is the current applied to the Hall bar, ${R_{PHE}}$ is the planar Hall resistance and ${\phi _M}$ is the angle of the magnetization with respect to the current direction. In the limit where $B \gg {B_A}$, ${\phi _M} = \phi  - ({{{B_A}} \mathord{\left/
 {\vphantom {{{B_A}} B}} \right.
 \kern-\nulldelimiterspace} B})\sin (2\phi  - 2{\phi _{EA}})$, where ${B_A}$ is the magnitude of the in-plane magnetic anisotropy in the Permalloy and ${\phi _{EA}}$ is the angle of the magnetic easy-axis with respect to the current direction. The first harmonic Hall measurement is used to determine ${I_0}{R_{PHE}}$, ${B_A}$, and ${\phi _{EA}}$.

\begin{figure*}[!t]
\centering
\includegraphics[width=14 cm]{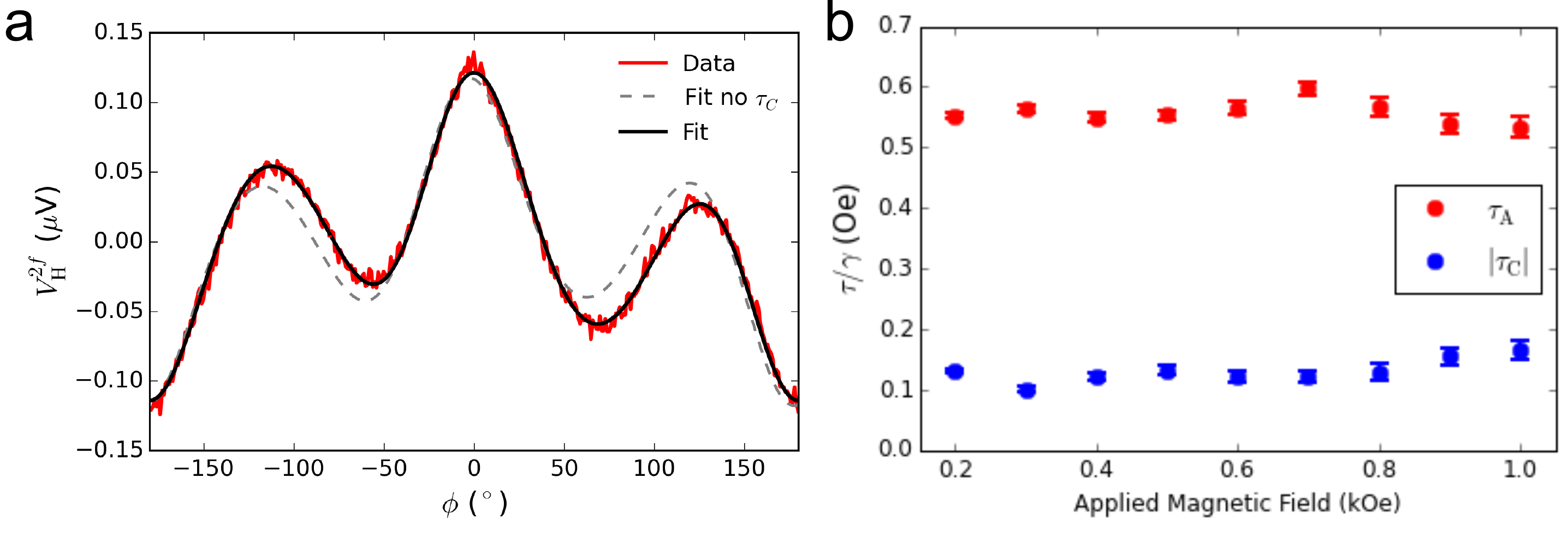}
    \caption[ Second harmonic Hall measurement of the torques in a TaTe$_2$/Py device]{(a) Second harmonic Hall voltage for Device 16, where the dashed grey line gives the fit function assuming only Rashba-like torque contributions and a small in-plane magnetic anisotropy, and the black curve shows the full fit using Eq.\ \ref{TaTe2SHapproxNernst}. The magnitude of the magnetic field is 20 mT and a 300 mV RMS excitation is applied across a series connection of the 480 $\Omega$ sample and a 50 $\Omega $ bias resistor. (b) The second harmonic Hall fit values of ${\tau _{\rm{A}}}$ and ${\tau _{\rm{C}}}$ as a function of applied in-plane magnetic field for Device 16 under the same excitation conditions. }
    \label{TaTe2secondharm}
\end{figure*}

$V_{\rm{H}}^{2f}$is related to the out-of-plane (${\tau _ \bot }$) and in-plane (${\tau _\parallel }$) components of the current-generated spin-orbit torques by\cite{Hayashi2014,Avci2014,MacNeill2017}:
\small
\begin{equation}
\begin{aligned}
	V_{\rm{H}}^{2f} \approx & {I_0}{R_{PHE}}\cos (2{\phi _M})\frac{{{{{\tau _ \bot }} \mathord{\left/
 {\vphantom {{{\tau _ \bot }} \gamma }} \right.
 \kern-\nulldelimiterspace} \gamma }}}{{B + {B_A}\cos (2{\phi _M} - 2{\phi _{EA}})}} \\
 &+ \frac{{{I_0}{R_{AHE}}}}{2}\frac{{{{{\tau _\parallel }} \mathord{\left/
 {\vphantom {{{\tau _\parallel }} \gamma }} \right.
 \kern-\nulldelimiterspace} \gamma }}}{{B + {\mu _o}{M_{{\rm{eff}}}} + {B_A}{{\cos }^2}({\phi _M} - {\phi _{EA}})}},
\end{aligned}
\label{TaTe2SHapprox}
\end{equation}
\normalsize
where  ${R_{PHE}}$ is the planar Hall resistance, ${R_{AHE}}$ is the anomalous Hall resistance, ${\mu _o}{M_{{\rm{eff}}}}$ is the effective magnetization field, and $\gamma $ is the gyromagnetic ratio. In high symmetry systems the spin-orbit torques have a purely Rashba-like spin-symmetry with an out-of-plane component ${\vec \tau _ \bot } \propto \hat m \times \hat y$ and an in-plane component ${\vec \tau _\parallel } \propto \hat m \times (\hat m \times \hat y)$ where the applied current is the $\hat x$ direction. In this case, both torque magnitudes are proportional to $\cos ({\phi _M})$ and therefore $V_{\rm{H}}^{2f}(\phi ) = V_{\rm{H}}^{2f}( - \phi )$ for small ${B_A}$.  Figure\ \ref{TaTe2secondharm}a shows $V_{\rm{H}}^{2f}$as a function of $\phi $ for one of our Hall bar devices (Device 16). As in our ST-FMR measurements, the measured second harmonic Hall signal clearly lacks $\phi  \to  - \phi $ symmetry. This asymmetry cannot be captured by Rashba-like torques and the small in-plane magnetic anisotropy (grey fit in Fig.\ \ref{TaTe2secondharm}a), pointing to the presence of additional torques.
If we allow for a Dresselhaus-like component of field-like torque, $ \propto \hat m \times \hat x$, and model the out-of-plane torques present in our Hall bar as a sum of Rashba-like and Dresselhaus-like components:
	\[{\tau _ \bot } = A\cos (\phi ) + C\sin (\phi ),\] 
 we can accurately capture the $\phi  \to  - \phi $ symmetry breaking in the observed $V_{\rm{H}}^{2f}$.

To fit the second harmonic Hall data, one must also consider the effects of magnetothermal voltages\cite{Avci2014}. For an in-plane magnetization, thermal contributions to the second harmonic Hall voltage are dominated by the planar Nernst effect arising from an out-of-plane thermal gradient, which adds a term proportional to $\cos (\phi )$ to $V_{\rm{H}}^{2f}$. In the limit of ${\mu _0}{M_{{\rm{eff}}}} \gg B,{B_A}$, the second harmonic Hall voltage arising from ${\tau _\parallel }$ also has an overall angular dependence of $\cos (\phi )$ if ${\tau _\parallel }$ has only a Rashba-like contribution. We therefore combine the terms proportional to $\cos (\phi )$ into one fit parameter to yield a total fit function of:
\begin{equation}
	V_{\rm{H}}^{2f} \approx \cos (2{\phi _M})\frac{{A\cos ({\phi _M}) + C\sin ({\phi _M})}}{{B + {B_A}\cos (2{\phi _M} - 2{\phi _E})}} + N\cos ({\phi _M}) + {\rm{offset}},\label{TaTe2SHapproxNernst}
\end{equation}
 	
where $A$ is proportional to the out-of-plane torques with a $\cos ({\phi _m})$ dependence,
	\[A = {I_0}{R_{PHE}}{{{\tau _{\rm{A}}}} \mathord{\left/
 {\vphantom {{{\tau _{\rm{A}}}} \gamma }} \right.
 \kern-\nulldelimiterspace} \gamma },\] 
and $C$ is proportional to the out-of-plane torques with a $\sin ({\phi _m})$ dependence, 
	\[C = {I_0}{R_{PHE}}{{{\tau _{\rm{C}}}} \mathord{\left/
 {\vphantom {{{\tau _{\rm{C}}}} \gamma }} \right.
 \kern-\nulldelimiterspace} \gamma }.\] 
$N$ is a combination of an in-plane antidamping torque and the planar Nernst contributions. Figure\ \ref{TaTe2secondharm}a shows the measured $V_{\rm{H}}^{2\omega }$ as a function of $\phi $ (red) for Device 16 and the fit (black). Figure\ \ref{TaTe2secondharm}b shows ${{{\tau _{\rm{A}}}} \mathord{\left/
 {\vphantom {{{\tau _{\rm{A}}}} \gamma }} \right.
 \kern-\nulldelimiterspace} \gamma }$ and ${{{\tau _{\rm{C}}}} \mathord{\left/
 {\vphantom {{{\tau _{\rm{C}}}} \gamma }} \right.
 \kern-\nulldelimiterspace} \gamma }$ as a function of applied magnetic field, where ${I_0}{R_{PHE}}$ extracted from the first harmonic Hall signal is divided from the fit values of $A$ and $C$. The field independence of the extracted terms confirms their origin as current-generated torques.

\begin{table*}[!tp]
\centering
\scalebox{0.85}{
    \begin{tabular}{|c| c| c| c| c| c| c| c| c|}
    \hline
     Device & Device Type & $t$ (nm)  & $L\times W$ ($\mu$m) & $C/A$&  $\tau_{\mathrm{S}}/\tau_{\mathrm{A}}$&$B_{\mathrm{A}}$ & $\phi_{I-ac}$& $\phi^{Raman}_{I-ac}$  \\ 
   Number   &  & $\pm$ 0.3 nm & $\pm$ 0.2 $\mu$m & &  &(0.1 mT)&  & $\pm 2^{\circ}$  \\ \hline
      
	1	& ST-FMR &8.8	&5 X 4&	-0.687(7)&	0.20(2)&	23&	29&	45 \\ \hline
	2	&ST-FMR	&19.7	&5 X 4&	0.384(7)&	-0.04(2)&	20&	-57&	-70 \\ \hline
	3	&ST-FMR	&6.8	&4 X 3&	-0.31(1)&	0.17(3)&	48&	70&	50 \\ \hline
	4	&ST-FMR	&11.3	&5 X 4&	-0.596(8)&	0.30(2)&	20&	21&	25 \\ \hline
	5	&ST-FMR	&10.4	&5 X 4&	-0.238(6)&	0.18(2)&	29&	70&	-- \\ \hline
	6	&ST-FMR	&9.1	&5 X 4&	-0.275(6)&	0.22(1)&	21&	63&	65 \\ \hline
	7	&ST-FMR	&15.4	&5 X 4&	0.197(6)&	0.21(2)&	28&	-7&	-7 \\ \hline
	8	&ST-FMR	&16.4	&5 X 4&	0.147(6)&	-0.02(2)&	16&	-86&	-90 \\ \hline
	9	&ST-FMR	&9.4	&3 X 2&	0.189(4)&	0.18(1)&	17&	-72&	-- \\ \hline
	10	&ST-FMR	&11.0	&4 X 3&	-0.521(5)&	0.18(1)&	25&	9&	-- \\ \hline
	11	&ST-FMR	&16.1	&4 X 3&	0.02(1)&	-0.01(4)&	53&	-90&	-90 \\ \hline
	12	&ST-FMR	&6		&4 X 3&	0.334(9)&	0.16(3)&	48&	-68&	-55 \\ \hline
	13	&ST-FMR	&8.2	&4 X 3&	-0.46(1)&	0.15(3)&	43&	69&	50 \\ \hline
	14	&ST-FMR	&17.4	&4 X 3&	-0.50(1)&	0.06(3)&	25&	56&	25 \\ \hline
	15	&ST-FMR	&4.5	&4 X 3&	-0.23(1)&	0.27(3)&	29&	73&	20 \\ \hline
	16	&SH	&14.2		&10.3 X 3&	0.237(6)&	--&		19&	-15&	-20 \\ \hline
	17	&SH&	7.8		&7.2 X 2&	-0.052(4)&		--&		31&	80&	85 \\ \hline
	18	&SH	&16.4		&4.9 X 2&	0.282(5)&		--&		18&	-59&	-20 \\ \hline
	19	&SH&	5.0		&16 X 3&	0.182(8)&		--&		46&	-71&	-- \\ \hline

    \end{tabular}}
    \caption{Comparison of device parameters, torque ratios, and magnetic anisotropy parameters for TaTe$_2$/Py heterostructures, where $\phi_{I-ac}=\phi_{EA}-90^\circ$.}\label{tab:TaTe2}

\end{table*}

\section{Cross-sectional HAADF-STEM}\label{STEMSec} 
For electron microscopy measurements, we prepare a thin cross-sectional lamella from the active area of Devices 7 and 11 using the focused ion beam (FIB) lift-out technique. Imaging is performed perpendicular to the current direction of the sample. Aberration-corrected high-angle annular dark-field (HAADF) STEM is performed in an FEI Titan Themis operating at 300 kV. The convergence semi-angle is 21.4 mrad, and the inner collection angle for HAADF is 68 mrad. The probe current is ~50-60 pA. To overcome drift and scan noise, we collect stacks of 30 images taken with 1 $\mu$s/pixel dwell time and align and average them using rigid registration. Despite the high voltage, we do not observe knock-on damage between frames or during imaging.

\section{Measurement of the in-plane magnetic anisotropy}\label{TaTe2anis}

\begin{figure*}[!t]
\centering
\includegraphics[width=14 cm]{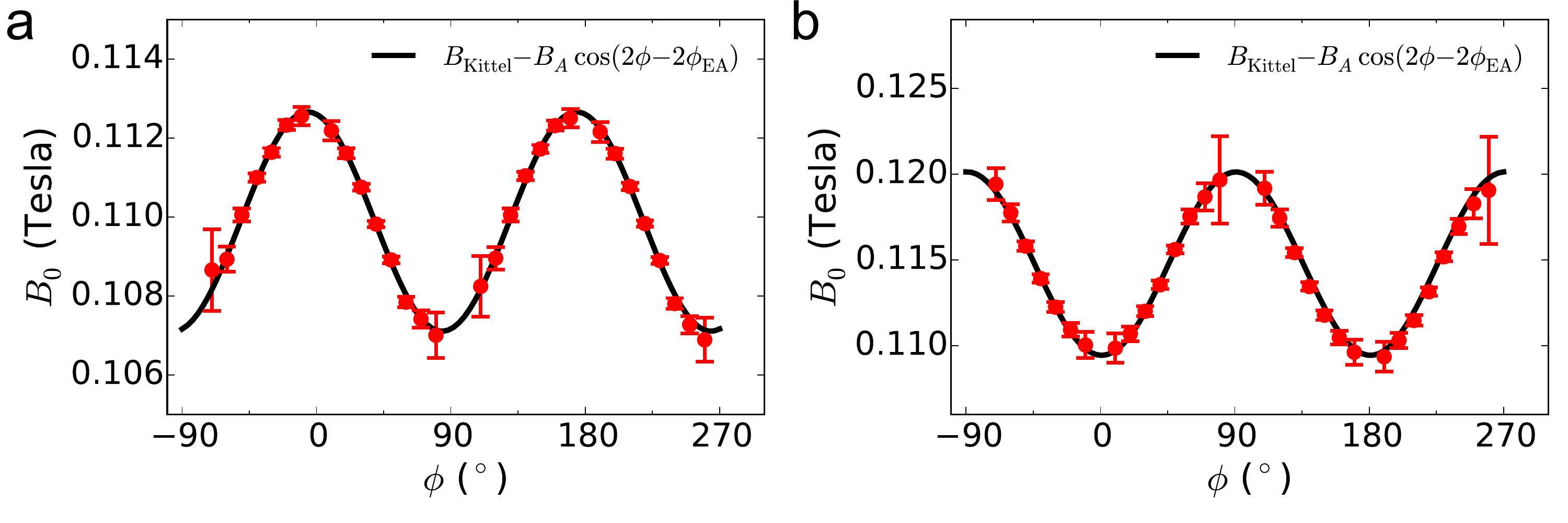}
    \caption[Resonant field for ST-FMR devices as a function of applied magnetic field angle.]{Resonant field for ST-FMR devices as a function of applied magnetic field angle for Device 7 (a) and Device 11 (b). The angle at which the resonant field is minimized gives the direction of the magnetic easy axis, ${\phi _{EA}}$, here $83^\circ $ and $0^\circ $, corresponding to ${\phi _{I - ac}}$ values of $ - 7^\circ $ and $ - 90^\circ$, respectively. The magnitude of the magnetic easy axis, ${B_A}$,  can also be directly extracted from the fit. The applied microwave frequency is 9 GHz with applied powers of 2 dBm and 5 dBm respectively. }
    \label{TaTe2fig7}
\end{figure*}

\begin{figure*}[!t]
\centering
\includegraphics[width=14 cm]{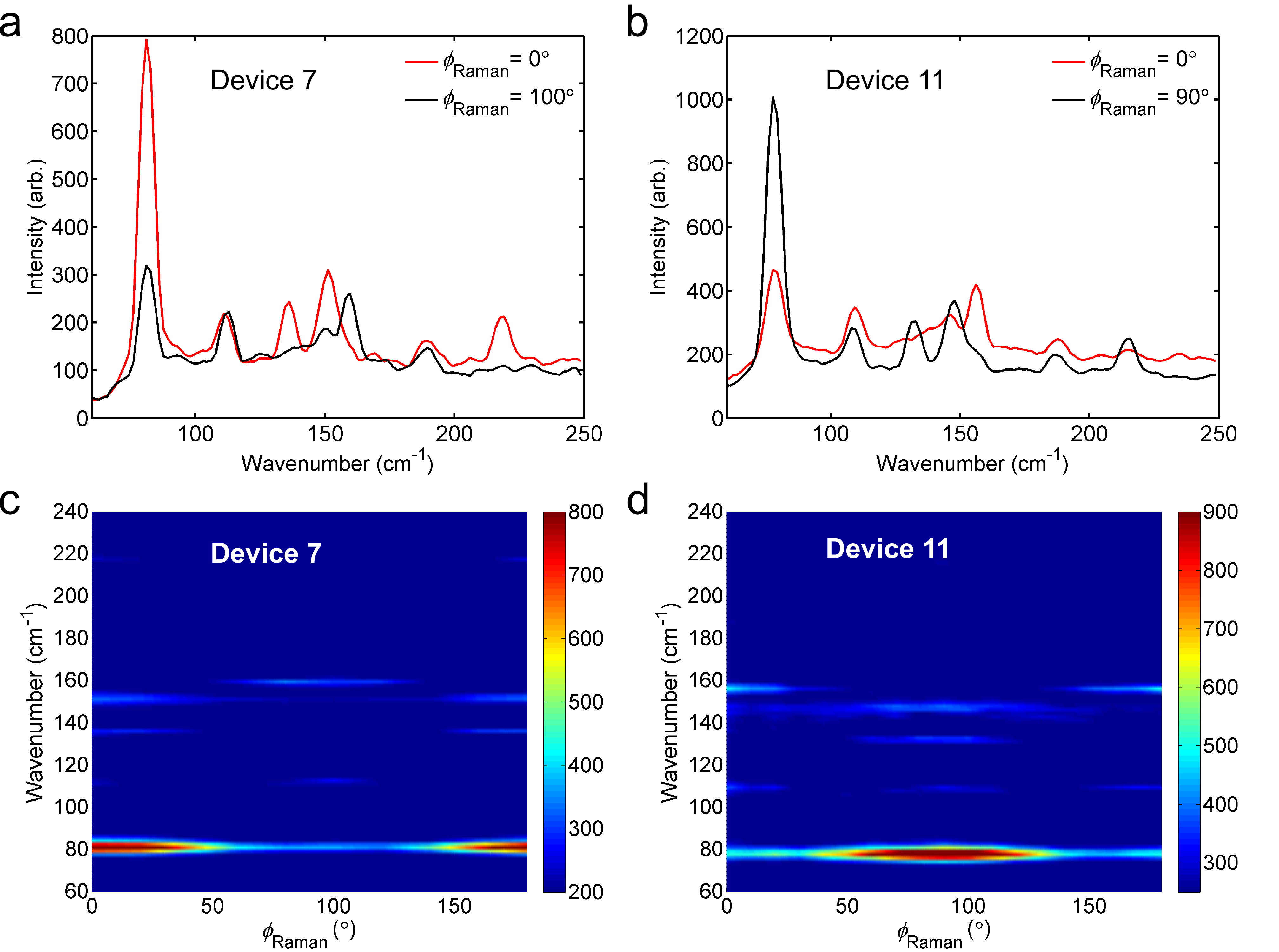}
    \caption[Polarized Raman spectroscopy for TaTe$_2$.]{(a,b) Raman spectra for Devices 7 and 11, with a 488 nm excitation and with the excitation and detector polarized parallel to each other. ${\phi _{{\rm{Raman}}}}$ is the angle between the excitation polarization and the device current direction (along the bar).  The red traces show spectra with the polarization parallel to the current and the black traces show spectra with the polarization approximately perpendicular. (c,d) Angular dependence of the Raman spectra for the two devices. The color map represents the peak intensity (with arbitrary units). The maximum of the $ \sim 80$ cm$^{-1}$ peak corresponds to the TaTe$_2$ a-c mirror plane, where ${\phi _{{\rm{Raman}}}} \to  - {\phi^{Raman} _{I - ac}}$.}
    \label{TaTe2fig8}
\end{figure*}

In our ST-FMR devices, measurements of the resonant field (${B_0}$) as a function of the applied magnetic field angle ($\phi $) can be used to extract the in-plane magnetic easy-axis direction and magnitude. The angular dependence of ${B_0}$ can be described by:
	\[{B_0} = {B_{{\rm{Kittel}}}} - {B_A}\cos (2[\phi  - {\phi _{EA}}]),\] 
where ${B_{{\rm{Kittel}}}}$ is the resonance field without any in-plane anisotropy and $\phi _{EA}$ is the angle of the easy-axis with respect to the current direction. Figure\ \ref{TaTe2fig7} shows the magnetic field at ferromagnetic resonance as a function of the in-plane magnetization angle for Devices 7 and 11. The data from both samples indicate the presence of a uniaxial magnetic anisotropy within the sample plane, at angles of 83 and 0 degrees from the current direction and correspond to ${\phi _{I - ac}}$ values of $ - 7^\circ $ and $ - 90^\circ $, respectively. Table\ \ref{tab:TaTe2} shows the magnitude of ${B_A}$ and $\phi_{I-ac}=\phi _{EA}-90^\circ$ for all devices.

In our Hall bar devices, measurements of the first harmonic Hall voltage (described in Supporting Information Section\ \ref{SecHarmHallSec}) can be used to determine the magnitude and direction of the induced easy-axis in the Permalloy film. The Hall voltage is given by ${V_{\rm{H}}} = {I_0}{R_{PHE}}\sin (2{\phi _M}),$ where ${I_0}$ is the current applied to the Hall bar, ${R_{PHE}}$ is the planar Hall resistance and ${\phi _M}$ is the angle of the magnetization with respect to the current direction. In the limit where $B \gg {B_A}$, ${\phi _M} = \phi  - ({{{B_A}} \mathord{\left/
 {\vphantom {{{B_A}} B}} \right.
 \kern-\nulldelimiterspace} B})\sin (2\phi  - 2{\phi _{EA}})$. The first harmonic Hall measurement is used to determine ${I_0}{R_{PHE}}$, ${B_A}$, and ${\phi _{EA}}$. Figure\ \ref{TaTe2firstharm} shows the first harmonic Hall voltage for Device 16. 
 
We find that TaTe$_2$ and WTe$_2$ both induce a magnetic easy axis in the adjacent Py, but in opposite directions with respect to their respective crystallographic mirror plane direction. TaTe$_2$ induces a magnetic easy-axis perpendicular to the a-c mirror plane, whereas WTe$_2$ induces a magnetic easy-axis along its b-c mirror plane. The magnitude of the induced easy axis is stronger in WTe$_2$/Py bilayers\cite{MacNeill2016,MacNeill2017} (6.2-17.3 mT) as compared to the TaTe$_2$/Py bilayers (1.7-5.3 mT).  

\begin{figure*}[!t]
\centering
\includegraphics[width=14 cm]{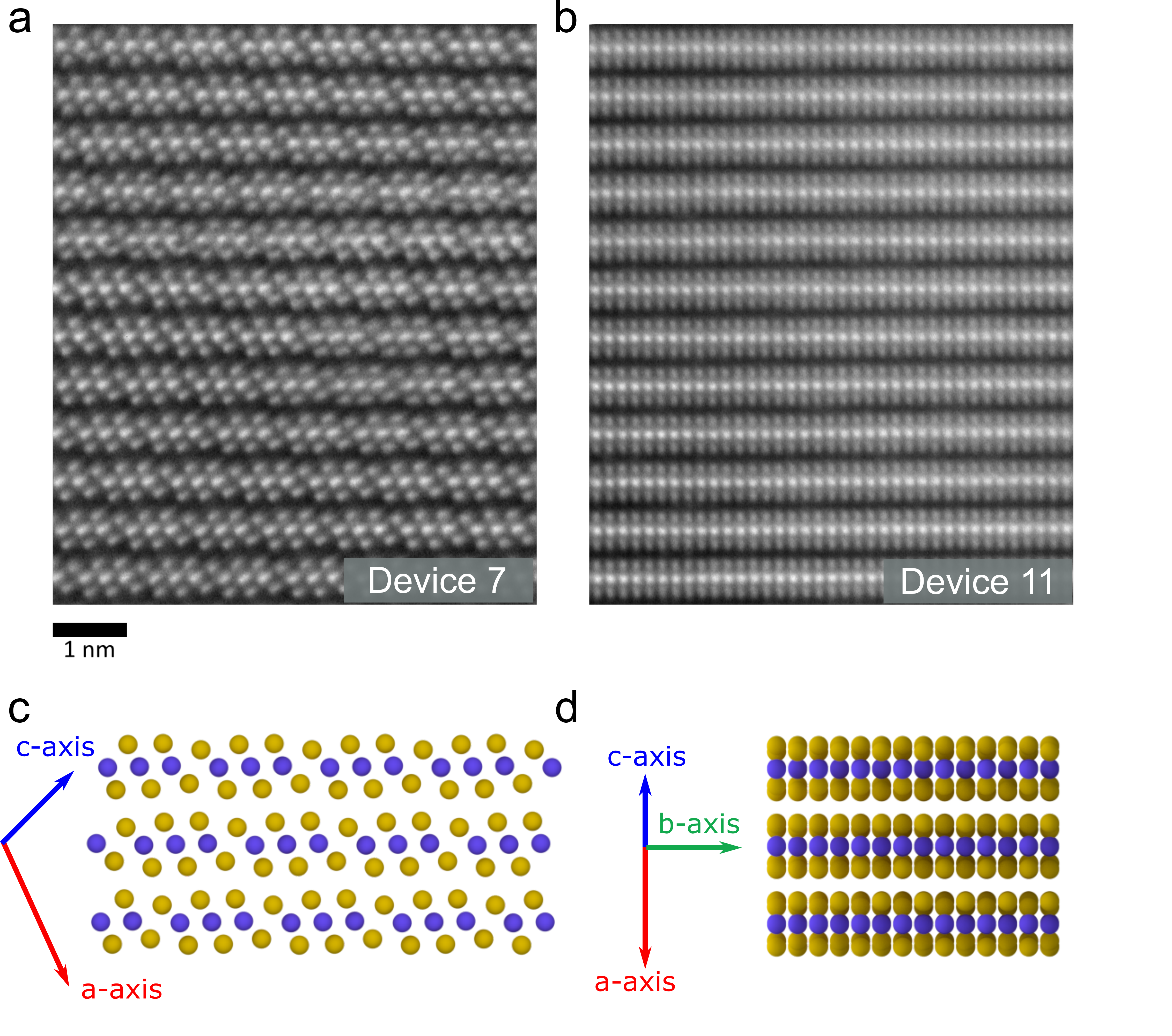}
    \caption[ Cross-sectional HAADF-STEM images along and perpendicular to the TaTe$_2$ mirror plane.]{(a,b) Cross-sectional HAADF-STEM images of Devices 7 and 11 respectively. The images are taken perpendicular to the current direction for each device (along the ST-FMR sample bar). (a) looks down the TaTe$_2$ b-axis and (b) perpendicular to the TaTe$_2$ b-axis, confirming that in each device the induced magnetic easy-axis lies along the TaTe$_2$ b-axis. (c) and (d) show the relative orientations of (a) and (b) respectively.}
    \label{TaTe2fig9}
\end{figure*}

\section{Determination of TaTe$_2$ crystal axes}\label{TaTe2RamanTEM}
As noted in the previous section, similar to WTe$_2$/Py bilayers, TaTe$_2$ induces a magnetic easy axis in the adjacent Permalloy layer. We use this easy axis to determine the angle between the current direction and the TaTe$_2$ crystal axes.  For TaTe$_2$/Py, we found that the alignment of the magnetic easy axis is along the b-axis of the TaTe$_2$ crystal using a combination of magnetic anisotropy measurements together with polarized Raman spectroscopy and cross-sectional HAADF-STEM imaging. Details of the HAADF-STEM imaging can be found in Supporting Information Section\ \ref{STEMSec}. The Raman measurements are performed using a Renishaw inVia confocal Raman microscope with a linearly polarized 488 nm wavelength excitation and a co-linear polarizer placed in front of the spectrometer entrance slit. The sample is positioned such that the excitation electric field is in the sample plane with a normal angle of incidence. 

No previous measurements of a polarization-dependent Raman spectrum in TaTe$_2$ have been reported to our knowledge, but the symmetry of the polarization dependence of the modes are required to be the same as the room temperature monoclinic phase in 1T'-MoTe$_2$ (space group \# 11), which has been studied in detail\ \cite{Beams2016ACSnano,ChenNanoLett2016,ZhouJACS2017,Wang2017}. This is because the polarization dependence of the Raman signal is governed by the point group (2/m) which is the same for both TaTe$_2$ and 1T'-MoTe$_2$. Raman spectra are shown for two samples (Devices 7 and 11) in the Fig.\ \ref{TaTe2fig8}. We have verified by cross-sectional HAADF-STEM imaging in both of these devices (Fig.\ \ref{TaTe2fig9}) that the $ \sim 80$ cm$^{-1}$ parallel-polarized Raman mode in TaTe$_2$ is maximized along the crystallographic a-c mirror plane and minimized along the b-axis. For each of these devices, the measured magnetic easy-axis corresponds to the TaTe$_2$ b-axis (Fig.\ \ref{TaTe2fig7}). The values of ${\phi _{I - ac}}$ as measured by Raman (${\phi^{Raman} _{I - ac}} \to  - {\phi _{Raman}}$) are reported in Table\ \ref{tab:TaTe2} for 15 of our devices. For 12 of these devices the induced magnetic easy axis lies within $20^\circ $ of the estimated peak of the Raman 80 cm$^{-1}$ mode (the a-c plane). For the remaining three devices there is significant disagreement. A shape anisotropy in the Py bar can account for some of the discrepancy. In our ST-FMR samples the etched bar length is $ \sim 40$ ${\rm{ }}\mu {\rm{m}}$ long (the majority of the bar is covered by the top leads) leading to an aspect ratio of 3:40 or 4:40 and a shape anisotropy of $ \sim 10$ Oe. In at least one of these samples (Device 18) the discrepancy may be explained by mild damage that occurred to the device between the electronic measurements and Raman characterization.

\section{In-plane torques in TaTe$_2$/Py bilayers}\label{SecIPT}

\begin{figure*}[!t]
\centering
\includegraphics[width=14 cm]{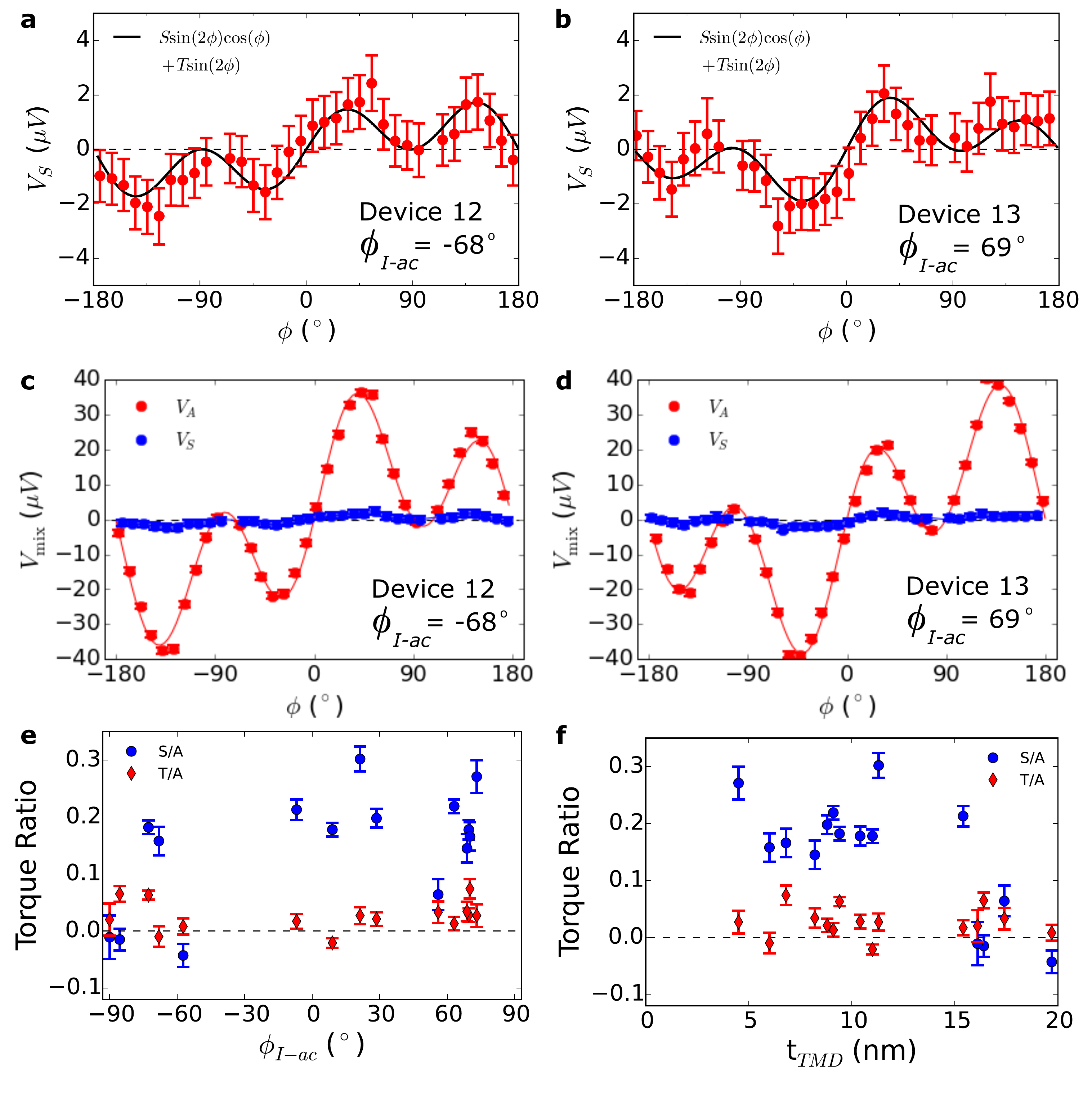}
    \caption[ In-plane torques in our TaTe$_2$/Py devices.]{(a,b) Symmetric component of the mixing voltage (red points) for ST-FMR devices 12 and 13, corresponding to the in-plane torque. The black curves show fits using Eq.\ \ref{TaTe2Vsangle}. (c,d) Symmetric (${V_S}$, blue) and antisymmetric (${V_A}$, red) components of the mixing voltage, corresponding to in-plane and out-of-plane torques, plotted on the same scale. (e,f) Ratios of the fit components $S/A$ and $T/A$ vs. ${\phi _{I - ac}}$ and ${t_{{\rm{TMD}}}}$. In (a-f) the applied microwave frequency is 9 GHz.}
    \label{TaTe2InPlaneTau}
\end{figure*}

In the main text we focused on the out-of-plane torques in TaTe$_2$/Py bilayers. Here, we will comment on the in-plane torques present in our samples as measured by ST-FMR. The symmetric component of the Lorentzian fit of the mixing voltage is proportional to the in-plane torques in the bilayers. We observe a small but non-zero torque component with an angular dependence proportional to $\cos (\phi )$ in our samples, corresponding to a torque $ \propto \hat m \times (\hat m \times \hat y)$, which we call ${\tau _{\rm{S}}}$. The angular dependence of the in-plane torques is shown for two different devices in Fig.\ \ref{TaTe2InPlaneTau}a and b, and the ratio of this torque to ${\tau _{\rm{A}}}$ is reported in Table\ \ref{tab:TaTe2} for all devices measured. The relatively small size of the measured symmetric signals (blue) is evident when plotted with the corresponding antisymmetric signals (red) on the same scale (Fig.\ \ref{TaTe2InPlaneTau}c and d). In some of our samples, we also observe a very small in-plane torque $ \propto \hat m \times \hat z$, which we call ${\tau _{\rm{T}}}$. We therefore fit the dependence of the applied magnetic field angle for the symmetric component of the ST-FMR mixing voltage, ${V_{\rm{S}}}$, as a sum of these two torques:	
\begin{equation}
	{V_S} = \sin (2\phi )\left[ {S\cos (\phi ) + T} \right].\label{TaTe2Vsangle}
\end{equation}
The torque ${\tau _{\rm{T}}}$ is allowed by the symmetry of the TaTe$_2$ crystal, however, this small component of torque is inconsistent across our devices, and does not show the expected dependence across the a-c mirror plane, $i.e.$ it does not always change sign for positive and negative ${\phi _{I - ac}}$ as expected by symmetry. We therefore do not report it as a torque ubiquitous to the TaTe$_2$/Py system. Figures\ \ref{TaTe2InPlaneTau}e and f show $S/A$ and $T/A$ as a function of ${\phi _{I - ac}}$ and the TMD thickness, ${t_{{\rm{TMD}}}}$ for all of our ST-FMR devices. 
Due to the small size of the in-plane torques in the TaTe$_2$/Py bilayer system, we do not report the in-plane torques as measured by the second harmonic Hall technique as it is difficult to separate such a small term from the Nernst voltage present in our samples. 

\section{Model for C/A}\label{CAmodelSec}

In the main text we have modeled the ${\phi _{I - ac}}$ dependence of the torques proportional to $\hat m \times \hat x$ as arising solely from a Dresselhaus symmetry and the torques proportional to $\hat m \times \hat y$ as a sum of Rashba and Dresselhaus symmetry contributions:
	\[\tau _ \bot ^C = C\sin (\phi ) = [D\sin (2{\phi _{I - ac}})]\sin (\phi ),\] 
	\[\tau _ \bot ^A = A\cos (\phi ) = [R + D\cos (2{\phi _{I - ac}})]\cos (\phi ),\] 
where $R$ is the component of $\cos (\phi )$ torques arising from a Rashba symmetry, and $D$ for Dresselhaus, which can generate both $\cos (\phi )$ and $\sin (\phi )$ dependent torques. The fit equation used to extract a ratio of ${D \mathord{\left/
 {\vphantom {D R}} \right.
 \kern-\nulldelimiterspace} R}$ from the plot of ${C \mathord{\left/
 {\vphantom {C A}} \right.
 \kern-\nulldelimiterspace} A}$  as a function of ${\phi _{I - ac}}$ in Fig. 3 of the main text is:
 \begin{equation}
 	\frac{C}{A} = \frac{{{D \mathord{\left/
 {\vphantom {D R}} \right.
 \kern-\nulldelimiterspace} R}\sin (2{\phi _{I - ac}})}}{{1 \pm {D \mathord{\left/
 {\vphantom {D R}} \right.
 \kern-\nulldelimiterspace} R}\cos (2{\phi _{I - ac}})}},\label{Eq:RandDmodel}
 \end{equation}
where the sign of ${D \mathord{\left/
 {\vphantom {D R}} \right.
 \kern-\nulldelimiterspace} R}$ sets the relative orientation of the field directions in the Dresselhaus and Rashba symmetries and the $ \pm $ in the denominator sets the overall orientation.  Our TaTe$_2$/Py bilayers are represented by the scenario shown in Fig. 1a and b of the main text, where ${D \mathord{\left/
 {\vphantom {D R}} \right.
 \kern-\nulldelimiterspace} R}$ is negative and the sign in the denominator is positive, consistent with an Oersted field from tilted currents dominating both contributions. Using this fit function, we extract a value of $D/R= -0.51\pm 0.03$. We note that this model does not take into account the possibility that $D$ and $R$ may differ in their dependence on the thickness of the TMD.

\section{Dresselhaus-like torque in WTe$_2$}\label{WTe2}

\begin{figure}[!t]
	\centering
	\includegraphics[width=13 cm]{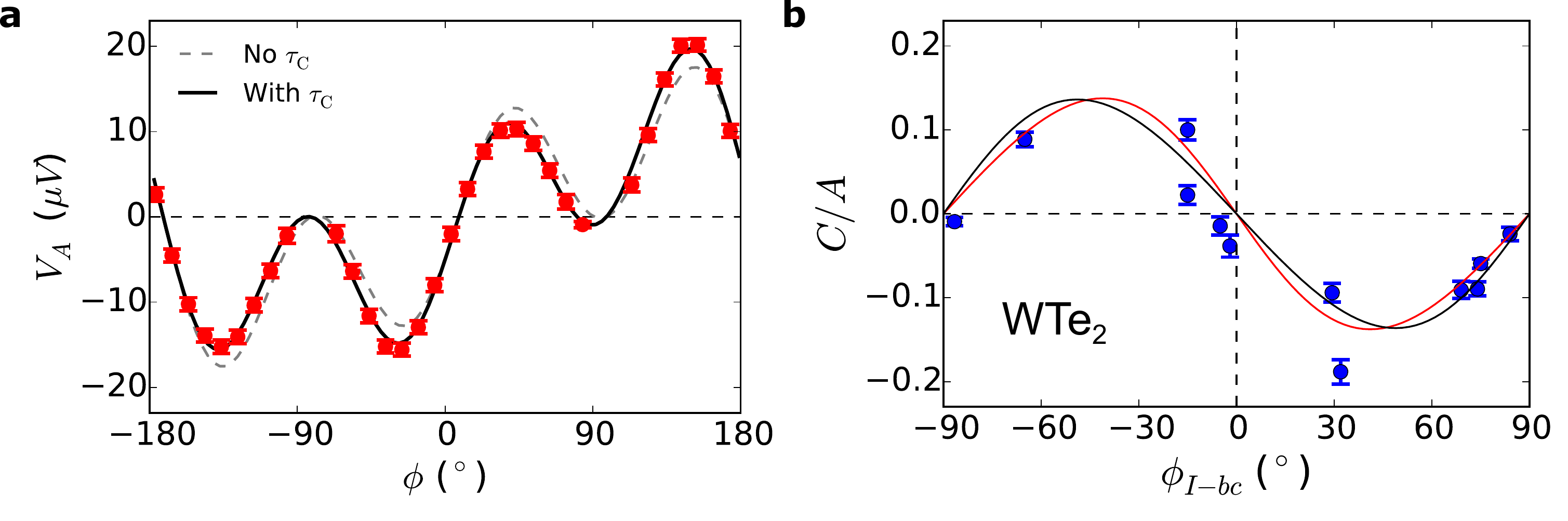}
	\caption{(a) Antisymmetric component of the ST-FMR resonance as a function of applied in-plane magnetic field angle for WTe$_2$/Py Device 11 in Table \ref{tab:WTe2Dress}. The applied microwave power is 2 dBm at a frequency of 9 GHz. The angle from the current to the mirror plane is 32$^{\mathrm{o}}$. (b) Ratio of torques $ \propto \hat m \times \hat x$  to the torques $ \propto \hat m \times \hat y$, C/A, as a function of the angle between the applied current and the WTe$_2$ b-c mirror plane for devices studied by ST-FMR (blue circles). Fits as discussed in Supporting Information Section \ref{WTe2}}
	\label{TaTe2fig4}
\end{figure}

Like TaTe$_2$, WTe$_2$ possesses a strong in-plane resistivity anisotropy, so one should expect it also to generate a field-like torque with Dresselhaus-like symmtry by the mechanism described in the main text.  In our previous work on WTe$_2$/Py devices, we did not originally notice a torque component with a Dresselhaus-like symmetry because we focused on samples with voltage applied perpendicular or parallel to the b-c mirror plane (${\phi _{a - I}} = 0^\circ $ and ${\phi _{a - I}} = 90^\circ $), where any Dresselhaus-like component must go to zero by symmetry.  Including a Dresselhaus-like component, $\hat m \times \hat x$, to the fit of the out-of-plane torques (${V_A}$, antisymmetric component of ${V_{mix}}$) for the WTe$_2$/Py ST-FMR devices previous studied\cite{MacNeill2016,MacNeill2017}, gives an overall dependence for the applied magnetic field angle, $\phi $, of:
	\[{V_A} = \sin (2\phi )[A\cos (\phi ) + B  + C\sin (\phi )],\] 	
where $A$ represents torques $ \propto \hat m \times \hat y$, $B$ represents torques $\propto \hat m \times (\hat m \times \hat z)$ and $C$ represents torques $ \propto \hat m \times \hat x$.  Figure\ \ref{TaTe2fig4}a shows data from a WTe$_2$/Py device with the voltage applied 32$^{\circ}$ from the b-c mirror plane, along with fits with and without the $C$ parameter. Table\ \ref{tab:WTe2Dress} shows the values of ${C \mathord{\left/
 {\vphantom {C A}} \right.
 \kern-\nulldelimiterspace} A}$ extracted from the fits from the devices from our previous study\cite{MacNeill2016,MacNeill2017}, as well as the values of ${B  \mathord{\left/
 {\vphantom {B  A}} \right.
 \kern-\nulldelimiterspace} A}$ determined with and without including the parameter $C$ in the fits. As can be seen from these data, torques with Dresselhaus symmetry are clearly distinguishable when the current is applied away from a high-symmetry crystal axis, and our previous observations of the out-of-plane antidamping torque in WTe$_2$/Py bilayers are not affect by the inclusion of this extra term. 
 
 \begin{table}[t!]
\centering
\scalebox{0.75}{
    \begin{tabular}{|c| c| c| c| c| c| c| c|}
    \hline
     Device& $t$ (nm)  & $L\times W$ ($\mu$m) & $\tau_{\mathrm{B}}/\tau_{\mathrm{A}}$&  $\tau_{\mathrm{B}}/\tau_{\mathrm{A}}$& $C/A$& $B_{\mathrm{A}}$ & $\phi_{\mathrm{I-bc}}$  \\ 
      Number&  $\pm$ 0.3 nm & $\pm$ 0.2 $\mu$m & w/out $C$ & w/ $C$  & & (0.1 mT) & $\pm 2^{\circ}$  \\ \hline

    1 & 5.5 & $4.8\times4$ & 0.373(4) & 0.372(3)& -0.010(5)	& 70.1(7) & -87 \\ \hline
    2 & 15.0 & $6\times4$ & 0.011(7) & 0.011(7)& -0.01(1)	& 151(2) &  -5  \\\hline
    5 & 8.2 & $6\times4$ & 0.133(8) & 0.132(8)& 0.10(1)	& 150(1) & -15 \\ \hline
   7 & 3.4 & $4\times3$ & 0.207(8) & 0.206(8)& 	0.02(1)	& 153(1) & -15 \\ \hline
   9 & 6.7 & $5\times4$ & 0.278(6) & 0.279(6)& 	0.089(9)	& 173(1) & -65 \\ \hline
	11 & 14.0 & $5\times4$ & -0.13(1) & -0.128(9)&-0.19(1) 	& 138(2) & 32 \\ \hline
   12 & 5.3 & $5\times4$ & -0.320(6) & -0.319(6)&-0.024(8) 	& 156(3) & 84 \\ \hline
    14 & 5.3 & $5\times4$ & 0.340(7) & 0.341(7)& -0.09(1)	& 140(1) & 69 \\ \hline
     15 & 5.5 & $5\times4$ & 0.332(7) & 0.331(6)& -0.060(6)	& 155(1) & 75 \\ \hline
    16 & 3.4 & $5\times4$ & 0.236(8) & 0.236(7)&-0.09(1) 	& 132(1) & 29 \\ \hline
    17 & 2.6 & $5\times4$ & 0.020(8) & 0.021(8)& -0.039(1)	& 20(2) & -2 \\ \hline
   18 & 5.0 & $5\times4$ & -0.451(7) & -0.444(6)&-0.090(8)	& 20(3) & 74 \\ \hline
 
    \end{tabular}}
    \caption{Comparison of device parameters, torque ratios, and magnetic anisotropy parameters for WTe$_2$/Py bilayers as originally detailed in Refs. [\citenum{MacNeill2016,MacNeill2017}], but with the addition of a Dresselhaus torque component to the fit for the out-of-plane torques. We find a small but nonzero value for the ratio $C/A$ for samples in which ${\phi _{I - bc}}$ is sufficiently different from 0$^{\circ}$ or $\pm90^{\circ}$. We also show that the inclusion of the Dresselhaus-like torque $ \propto \hat m \times \hat x$ does not affect our previously reported values for the out-of-plane antidamping torque component because the ratio $B/A$ is unchanged within measurement uncertainty.}\label{tab:WTe2Dress}

\end{table}
 
Figure\ \ref{TaTe2fig4}b shows the extracted values of ${C \mathord{\left/
 {\vphantom {C A}} \right.
 \kern-\nulldelimiterspace} A}$ plotted as a function of ${\phi _{I - bc}}$, the angle of the applied current to the WTe$_2$ b-c mirror plane (${\phi _{I - bc}} = {\phi _{a - I}} - 90^\circ $). The dependence of ${C \mathord{\left/
{\vphantom {C A}} \right.
 \kern-\nulldelimiterspace} A}$ on the direction of current with respect to the WTe$_2$ crystal axes is similar to our TaTe$_2$/Py samples: ${C \mathord{\left/
 {\vphantom {C A}} \right.
 \kern-\nulldelimiterspace} A}$ goes to zero when current is parallel or perpendicular to a mirror plane $({\phi _{I - bc}} = 0^\circ $ and $90^\circ $), and is maximal when current is applied between these two values. 
We model the dependence of ${C \mathord{\left/
 {\vphantom {C A}} \right.
 \kern-\nulldelimiterspace} A}$ using Eq.\ \ref{Eq:RandDmodel}, and extract a value of ${D \mathord{\left/
 {\vphantom {D R}} \right.
 \kern-\nulldelimiterspace} R} =  - 0.13 \pm 0.02$. We note that for our WTe$_2$/Py samples, we do not have sufficient resolution to accurately determine the sign in the denominator of Eq.\ \ref{Eq:RandDmodel}, with both giving the same ratio of ${D \mathord{\left/
 {\vphantom {D R}} \right.
 \kern-\nulldelimiterspace} R}$ within the fit error and with comparable residuals ($ + $, red curve; $ - $, black curve). Since we know from our previous work\cite{MacNeill2016,MacNeill2017,Guimaraes2018Nano} that the Rashba component of the out-of-plane torque is dominated by the Oersted field, we suggest that the sign in the denominator should be positive.
 
Using the two-point sheet resistance of our WTe$_2$/Py devices, we have extracted an in-plane resistivity anisotropy of $\mathtt{\sim}2$ in WTe$_2$, with the a-axis being less resistive. We find a value of $\rho_{\mathrm{a-axis}}=530\pm140$ $\mu\Omega$cm and a value of $\rho_{\mathrm{b-axis}}=1160\pm100$ $\mu\Omega$cm averaged across devices that have been exfoliated in nitrogen and vacuum. We note that this difference in surface treatment may affect the absolute values of the resistivities, but it would be surprising if it affected the sign of the anisotropy. The resistivity anisotropy for WTe$_2$ found here is consistent with both TaTe$_2$ and 1T'-MoTe$_2$\ \cite{HughesJPC1978} in that the metal-atom chain is the low resistance axis for all of these materials. This implies that qualitatively the torques with Dresselhaus symmetry in WTe$_2$/Py heterostructures can be described by the tilted currents induced through the resistivity anisotropy.

\section{Thickness dependence of the torques in TaTe$_2$}\label{TaTe2ThickdepSec}

\begin{figure*}[!t]
\centering
\includegraphics[width=14 cm]{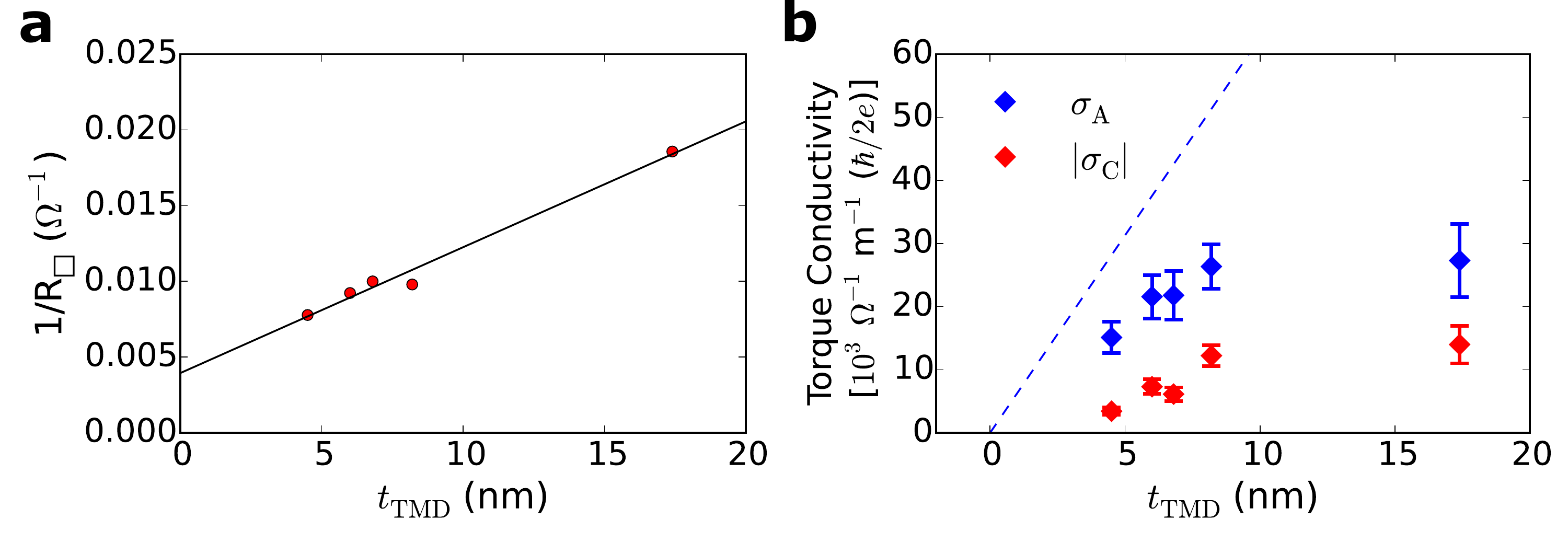}
    \caption{(a) $1/R_{\Box}$ as a function of the TaTe$_2$ thickness for 5 devices which have the same aspect ratio ($4\times 3$ $\mu$m) and are within $ \pm 10^\circ $ of $\left| {{\phi _{I - ac}}} \right| = 65^\circ $ (red circles). The fit (black line) allows an extraction of the resistivities for TaTe$_2$ and Py.  (b) Torque conductivities for the same 5 devices. Here, the dashed blue line gives an estimated Oersted field contribution to ${\sigma _A}$.
}
    \label{TaTe2ThicknessDep}
\end{figure*}

A torque conductivity is defined as the angular momentum absorbed by the magnet per second per unit interface area per unit electric field. It provides a measure of the torques produced in a spin source/ferromagnet bilayer independent of geometric factors. For a torque ${\tau _{\rm{K}}}$ (where K = A or C) we calculate the corresponding torque conductivity as:
	\[{\sigma _{\rm{K}}} = \frac{{{M_{\rm{S}}}lw{t_{{\rm{magnet}}}}}}{\gamma }\frac{{{\tau _{\rm{K}}}}}{{\left( {lw} \right)E}} = \frac{{{M_{\rm{S}}}l{t_{{\rm{magnet}}}}}}{\gamma }\frac{{{\tau _{\rm{K}}}}}{{{I_{{\rm{RF}}}} \cdot Z}},\] 
where ${M_{\rm{S}}}$ is the saturation magnetization, $E$ is the electric field, $l$ and $w$ are the length and width of the TaTe$_2$/Permalloy bilayer, $Z$ is the measured RF device impedance, $I_{\rm{RF}}$ is the RF current flowing in the device and ${t_{{\rm{magnet}}}}$ is the thickness of the Permalloy. We approximate ${\mu _0}{M_{\rm{S}}} \approx {\mu _0}{M_{eff}}$ = 0.83 T, as extracted from fits to the frequency dependence of the ST-FMR resonance field. Values of ${\tau _{\rm{K}}}$ are extracted from Eqs. 1 and 2 of the main text as described in Ref. \citenum{MacNeill2016}. $I_{\rm{RF}}$ and $Z$ are determined by measurements of the RF device and circuit parameters S11 and S21 with a vector network analyzer.

Here we report the thickness dependence of torque conductivities ${\sigma _A}$ and ${\sigma _C}$ for 5 devices $ \pm 10^\circ $ within $\left| {{\phi _{I - ac}}} \right| = 65^\circ $ where all selected devices also have the same aspect ratio of $4\times 3$ $\mu$m.  In Fig.\ \ref{TaTe2ThicknessDep}a we plot $1/\mathrm{R}_{\Box}$ for these 5 devices as a function of TaTe$_2$ thickness, where $\mathrm{R}_{\Box}=w\mathrm{R}/l$ is the sheet resistance as determined from the two point resistance of the device, ${\rm{R}}$, with width $w$ and length $l$. The fit function is given by:
	\[\frac{1}{{{{\rm{R}}_{{\Box}}}}} = \frac{{{t_{{\rm{mag}}}}}}{{{\rho _{\rm{residual}}}}} + \frac{{{t_{{\rm{TMD}}}}}}{{{\rho _{{\rm{TMD}}}}}}\] 
where ${t_{{\rm{mag}}}}$ is the thickness of the Py, $\rho _{\mathrm{residual}}$ is the residual resistivity at ${t_{{\rm{TMD}}}}=0$ and includes the Py resistance and device contact resistance, and ${\rho _{{\rm{TMD}}}}$ is the resistivity of TaTe$_2$ for $\left| {{\phi _{I - ac}}} \right| = 65^\circ $. We extract values of ${\rho _{{\rm{TMD}}}} \sim 120 \pm 10{\rm{ }}$ $\mu \Omega {\rm{cm}}$ and $\rho _{\mathrm{residual}} \sim 150 \pm 20$ $\mu \Omega {\rm{cm}}$.
	Using these values, we can estimate the Oersted field contribution to the torque conductivities. For the torque $ \propto \hat m \times \hat y$ the estimated Oersted field torque conductivity is given by:
	\[{\sigma _{{\rm{Oe}}}} = \left( {\frac{e}{\hbar }} \right){\mu _0}{M_{\rm{S}}}{t_{{\rm{mag}}}}\sigma _{{\rm{TMD}}}^{{\rm{Charge}}}{t_{{\rm{TMD}}}},\] 
where we estimate ${M_{\rm{s}}} \approx {M_{{\rm{eff}}}}$ from the ST-FMR measurements and $\sigma _{{\rm{TMD}}}^{{\rm{Charge}}} \sim 8.3 \pm 0.5 \times {10^5}{\rm{ (}}{\Omega ^{ - 1}}{{\rm{m}}^{ - 1}})$ from the extracted resistivity. The blue line in Fig.\ \ref{TaTe2ThicknessDep}b gives the estimated ${\sigma _{{\rm{Oe}}}}$ contribution for ${\sigma _A}$. Note that this estimate is for a material with an isotropic resistivity and may overestimate the value as it does not capture the effects of tilted current paths, as well as any thickness dependence to the TaTe$_2$ resistivity. Also, as noted in the main text, if the resistivity of the Py is not uniform across its thickness (for instance due to increased scattering near the TaTe$_2$/Py interface) this could have the effect of decreasing the measured value Oersted torque $\propto\hat{m}\times\hat{y}$ due to competing Oersted torques from the TaTe$_2$ and current above the midline of the Py thickness. Fig.\ \ref{TaTe2ThicknessDep}b also shows the thickness dependence of $\sigma_C$. If the mechanism driving the generation of the Dresselhaus-like torque in our TaTe$_2$/Py samples was interfacial in origin, we would expect no dependence on the TMD thickness. Instead, $\sigma_C$ mimics the thickness dependence of $\sigma_A$, and increases dramatically in the 5 to 10 nm regime.

\bibliography{TaTe2}

\end{document}